%% file: main.tex
\documentclass[11pt]{article}
\usepackage{times}
\usepackage{a4wide}
\usepackage{latexsym}
\usepackage{amsmath}
\usepackage{amssymb}
\usepackage{ifthen}
\usepackage{mathrsfs}
\usepackage{graphics}
\usepackage{color}
\usepackage{stmaryrd}
\usepackage{amsthm}

\usepackage{calc}
\usepackage[dvipsnames]{xcolor}
\usepackage{tikz}
\usepackage{tikz-cd}
\usetikzlibrary{calc,shapes,arrows}

\usetikzlibrary{arrows,positioning}
\tikzset{
    >=stealth',
    punkt/.style={
           rectangle,
           rounded corners,
           draw=black, very thick,
           text width=6.5em,
           minimum height=2em,
           text centered},
    pil/.style={
           ->,
           thick,
           shorten <=2pt,
           shorten >=2pt,}
}

\newcommand{\nc}{\newcommand}
\nc{\rnc}{\renewcommand} \nc{\nev}{\newenvironment}
\rnc{\subsection}{\secdef\ssa\ssb}
\nc{\ssa}[2][default]{\par\vspace{1ex}\refstepcounter{subsection}\noindent\textbf{\thesubsection.
#1. }} \nc{\ssb}[1]{\par\vspace{2ex}\noindent\textbf{#1. }}

\rnc{\subsubsection}{\secdef\sssa\sssb}
\nc{\sssa}[2][default]{\par\vspace{1ex}\refstepcounter{subsubsection}\noindent\textit{\thesubsubsection.
#1. }} \nc{\sssb}[1]{\par\vspace{1ex}\noindent\textit{#1. }}

\makeatletter
\rnc{\@seccntformat}[1]{{\normalfont\bfseries{\csname
the#1\endcsname}\hspace{1pt}.\hspace{0.4em}}}
\rnc{\section}{\@startsection
        {section}%
        {1}%
        {0mm}%
        {-\baselineskip}%
        {0.5\baselineskip}%
        {\normalfont\normalsize\bfseries\centering}%
}
\renewcommand{\@makecaption}[2]{\begin{center}#1. #2\end{center}}
\makeatother

\newtheorem{theo}{Theorem}[section]
\newtheorem{lem}[theo]{Lemma}
\newtheorem{cor}[theo]{Corollary}
\newtheorem{prop}[theo]{Proposition}

\theoremstyle{definition}
\newtheorem{defn}[theo]{Definition}
\newtheorem{rem}[theo]{Remark}

\rnc{\proof}[1][{}]{\smallskip\noindent\textit{Proof #1: }}
\nc{\proofend}{\hfill$\Box$\vspace{\topsep}\par}

\nc{\bende}{\eqno$\Box$}
\nc{\benda}{\tag*{$\Box$}}
\nc{\claimbenda}{\tag*{$\dashv$}}

\rnc{\labelenumi}{(\arabic{enumi})}
\rnc{\labelitemi}{\text{--}}
\rnc{\phi}{\varphi}
\rnc{\epsilon}{\varepsilon}
\nc{\bigmid}{\;\big|\;}
\nc{\Bigmid}{\;\Big|\;}
\rnc{\max}{\textup{max}}
\rnc{\min}{\textup{min}}
\rnc{\log}{\textup{log}\;}

\nc{\FO}{\textup{FO}}

\nc{\str}[1]{\ensuremath{\mathcal #1}}
\nc{\cls}[1]{\ensuremath{\mathscr #1}}
\nc{\Str}{\textsc{Str}}

\nc{\ind}{\textup{ind}}
\nc{\ord}{\textup{ord}}
\nc{\univ}{\textup{uni}}
\nc{\comp}{\textup{comp}}
\nc{\Mod}{\textsc{Mod}}
\nc{\Graph}{\textsc{Graph}}
\nc{\fin}{\textup{fin}}
\nc{\Forb}{\textsc{Forb}}

\nc{\LT}{\text{\L o\'s-Tarski}}

\nc{\rand}[1]{\marginpar{\raggedright\footnotesize #1}}
\nc{\yrand}[1]{\rand{\textbf{Y: }#1}}
\nc{\jrand}[1]{\rand{\textbf{J: }#1}}

\begin{document}

\title{Forbidden Induced Subgraphs and the \LT\ Theorem}

\author{Yijia Chen\\\normalsize School of Computer Science\\
\normalsize Fudan University\\
\normalsize yijiachen@fudan.edu.cn\\
\and
J\"{o}rg Flum\\\normalsize Mathematisches Institut \\
\normalsize Universit\"{a}t Freiburg\\
\normalsize joerg.flum@math.uni-freiburg.de}

\date{}
\maketitle

\begin{abstract}
\noindent Let $\cls C$ be a class of finite and infinite graphs that is
closed under induced subgraphs. The well-known \LT\ Theorem from classical
model theory implies that $\cls C$ is definable in first-order logic (\FO)
by a sentence $\varphi$ if and only if $\cls C$ has a finite set of
forbidden induced finite subgraphs. It provides a powerful tool to show
nontrivial characterizations of graphs of small vertex cover, of bounded
tree-depth, of bounded shrub-depth, etc. in terms of forbidden induced
finite subgraphs. Furthermore, by the Completeness Theorem, we can compute
from $\varphi$ the corresponding forbidden induced subgraphs. We show that
this machinery fails on finite graphs.
\begin{itemize}
\item There is a class $\cls C$ of finite graphs which is definable in
    \FO\ and closed under induced subgraphs but has no finite set of
    forbidden induced subgraphs.

\item Even if we only consider classes $\cls C$ of finite graphs which
    can be characterized by a finite set of forbidden induced subgraphs,
    such a characterization cannot be computed from an \FO-sentence
    $\varphi$, which defines $\cls C$, and the size of the
    characterization cannot be bounded by $f(|\varphi|)$ for any
    computable function $f$.
\end{itemize}
\noindent Besides their importance in graph theory, the above results also
significantly strengthen similar known results for arbitrary structures.

\end{abstract}

\section{Introduction}

Many classes of graphs can be defined by a finite set of forbidden induced
finite subgraphs. One of the simplest examples is the class of graphs of
bounded degree. Let $d\ge 1$ and $\cls F_d$ consist of all graphs with vertex
set $\{1, \ldots, d+2\}$ and maximum degree exactly $d+1$. Then a graph $G$
has degree at most $d$ if and only if no graph in $\cls F_d$ is isomorphic to
an induced subgraph of $G$. Less trivial examples include graphs of small
vertex cover (attributed to Lov\'asz~\cite{fel19}), of bounded
tree-depth~\cite{din92}, and of bounded shrub-depth~\cite{ghnoor12}. As a
matter of fact, understanding forbidden induced subgraphs for those graph
classes is an important question in structural graph
theory~\cite{dovgiathi12,zas17,ganhlines19,gajkre20}. However, a
straightforward adaptation of a result in~\cite{fellan94} shows that it is in
general impossible to compute the forbidden induced subgraphs from a
description of classes of graphs by Turing machines.

It is folklore~\cite{atsdawgro08,sanadscha16} that characterization by
finitely many forbidden induced finite subgraphs is equivalent to
definability by a universal sentence of first-order logic (\FO). But only
very recently, it was realized~\cite{cheflu20} that such a characterization
can be further understood by the \LT\ theorem. \L o\'s~\cite{los55} and
Tarski~\cite{tar54} proved the first so-called preservation theorem of
classical model theory. In its simplest form it says that the class
$\Graph(\varphi)$ of finite and infinite graphs that are models of a
sentence~$\varphi$ of first-order logic is closed under induced subgraphs
(or, that $\varphi$ is preserved under induced subgraphs) if and only if
there is a universal \FO-sentence $\mu$ with $\Graph(\varphi)= \Graph(\mu)$.
Recall that a universal sentence~$\mu$ is a sentence of the form $\forall x_1
\ldots \forall x_k\, \mu_0$, where $\mu_0$ is quantifier-free.

Such a universal sentence $\mu= \forall x_1\ldots \forall x_k\, \mu_0$
expresses that certain patterns of induced subgraphs with at most $k$
vertices are forbidden. In fact, let $\cls F$ be a finite set of finite
graphs and denote by $\Forb(\cls F)$ the class of (finite and infinite)
graphs that do not contain an induced subgraph isomorphic to a graph in $\cls
F$. Then for a universal sentence $\mu$ as above we have
\begin{equation}\label{eq:Fkmu}
\Graph(\mu)= \Forb\big(\cls F_k(\mu)\big).
\end{equation}
Here for any \FO-sentence $\varphi$ and $k\ge 1$ by $\cls F_k(\varphi)$ we
denote the class of graphs that are models of~$\neg \varphi$ and whose
universe is $\{1, \ldots, \ell\}$ for some $\ell$ with $1\le \ell\le k$.
Clearly, $\cls F_k(\varphi)$ is finite.

We say that a class $\cls C$ of finite and infinite graphs is \emph{definable
by a finite set of forbidden induced subgraphs} if there is a finite set
$\cls F$ of finite graphs such that $\cls C= \Forb(\cls F)$. Hence the
graph-theoretic version of the \LT\ Theorem can be restated in the form:
\begin{tabbing}
(I)  \ \ \=
Let $\cls C$ be a class of finite and infinite graphs.
The following are equivalent: \\[1mm]
 \> \ \ (i) \ $\cls C$ is closed under induced subgraphs and \FO-axiomatizable. \\
 \> \ (ii) \ $\cls C$ is axiomatizable by a universal sentence. \\
 \> (iii) \ $\cls C$ is definable by a finite set of forbidden induced subgraphs.
\end{tabbing}
This version of the \LT\ Theorem is already contained, at least implicitly,
in the article~\cite{vau54} of Vaught published in~1954. In addition, it is
easy to see that the equivalence between (ii) and (iii) holds for any class
of finite graphs too.

\medskip
Note that we have repeatedly mentioned that in the \LT\ Theorem graphs are
allowed to be infinite. This is not merely a technicality.
In~\cite{cheflu20}, to obtain the forbidden induced subgraph characterization
of graphs of bounded shrub-depth using the \LT\ Theorem, one simple but vital
step is to extend the notion of shrub-depth to infinite graphs. Indeed,
Tait~\cite{tai59} exhibited a class $\cls C$ of finite structures (which
might be understood as colored directed graphs) which is closed under induced
substructures and \FO-axiomatizable. Yet, $\cls C$ is not definable by any
universal sentence, thus cannot be characterized by a finite set of forbidden
induced substructures. As the first result of this paper, we strengthen
Tait's result to graphs.

\begin{theo}\label{thm:Taitgraphs}
There is a class $\cls C$ of finite graphs with the following properties.
\begin{enumerate}
\item[(i)] $\cls C$ is closed under induced subgraphs and
    \FO-axiomatizable,

\item[(ii)] $\cls C$ is not definable by a finite set of forbidden
    induced subgraphs.
\end{enumerate}
\end{theo}

\noindent Even though we are interested in structural and algorithmic results
for classes of finite graphs, we see that in order to apply the \LT\ Theorem
for such purposes we have to consider classes of finite and infinite graphs.
So in this paper ``graph'' means finite or infinite graph. As in the
preceding result we mention it explicitly if we only consider finite graphs.

Complementing Theorem~\ref{thm:Taitgraphs} we show that it is even
undecidable whether a given \FO-definable class of finite graphs which is
closed under induced subgraphs can be characterized by a finite set of
forbidden induced subgraphs. More precisely:
\begin{theo}\label{thm:decisionfinuniv}
There is no algorithm that for any \FO-sentence $\varphi$ such that
\[
\Graph_{\fin}(\varphi):=
 \big\{G \bigmid \text{$G$ is a finite graph and a model of $\varphi$}\big\}
\]
is closed under induced subgraphs decides whether $\varphi$ is equivalent to
a universal sentence on finite graphs.
\end{theo}

\medskip
As mentioned at the beginning, for a class of finite graphs definable by a
finite set of forbidden induced subgraphs, it is preferable to have an
explicit construction of those graphs. This however turns out to be difficult
for many natural classes of graphs. For example, the forbidden induced
subgraphs are only known for tree-depth at most $3$~\cite{dovgiathi12}. Let
us consider the $k$-vertex cover problem for a constant $k\ge 1$. It asks
whether a given graph has a vertex cover (i.e., a set of vertices that
contains at least one endpoint of every edge) of size at most $k$. The class
of all \textsc{yes}-instances of this problem, finite and infinite, is closed
under induced subgraphs and \FO-axiomatizable by the \FO-sentence
\[
\varphi^k_{\textup{VC}} :=\varphi_{\Graph}\wedge
 \exists x_1\ldots \exists x_k\forall y\forall z
  \Big(Eyz\to\bigvee_{1\le \ell\le k}(x_\ell=y\vee x_\ell=z)\Big),
\]
where $\varphi_{\Graph}$ axiomatizes the class of graphs. Hence, by (I) the
class of $\textsc{yes}$-instances can be defined by a finite set of forbidden
 induced subgraphs. As the reader will notice it is by no means trivial
to find a universal sentence equivalent to $\varphi^k_{\textup{VC}}$. But on
the other hand, by the Completeness Theorem, we can search for such a
universal sentence by enumerating all possible universal sentences $\mu$ and
all possible proofs for $\vdash \varphi^k_{\textup{VC}}\leftrightarrow \mu$,
and then extract the corresponding forbidden induced subgraphs from $\mu$ as
in~\eqref{eq:Fkmu}.

To explain the hardness of constructing forbidden induced subgraphs, we prove
two negative results.

\begin{theo}\label{thm:finiteuniversalgraph}
There is no algorithm that for any \FO-sentence $\varphi$ which is equivalent
to a universal sentence $\mu$ on finite graphs computes such a $\mu$.

Or equivalently, there is no algorithm that for any \FO-sentence $\varphi$
such that
\[
\Graph_{\fin}(\varphi)= \Forb_{\fin}(\cls F)
\]
for a finite set $\cls F$ of graphs computes such an $\cls F$. Here,
\[
\Forb_{\fin}(\cls F)
 := \big\{G \bigmid \text{$G$ is a finite graph without
          induced subgraph isomorphic to a graph in $\cls F$}\big\}.
\]
\end{theo}

\medskip
\begin{theo}\label{thm:igurgraph}
Let $f: \mathbb N\to \mathbb N$ be a computable function. Then there is a
class $\cls C$ of finite graphs and an \FO-sentence $\varphi$ such that
\begin{enumerate}
\item[(i)] $\cls C= \Graph_{\fin}(\varphi)$.

\item[(ii)] $\cls C= \Graph_{\fin}(\mu)$ for \emph{some} universal sentence
    $\mu$, in particular $\cls C$ is closed under induced subgraphs.

\item[(iii)] For \emph{every} universal sentence $\mu$ with $\cls C=
    \Graph_{\fin}(\mu)$ we have $|\mu|\ge f(|\varphi|)$.
\end{enumerate}
\end{theo}

\medskip
Theorem~\ref{thm:finiteuniversalgraph} significantly strengthens the
aforementioned result of~\cite{fellan94}. Even if a class $\cls C$ of finite
graphs definable by a finite set of forbidden induced subgraphs is given by
an \FO-sentence $\varphi$ with $\cls C= \Graph_{\fin}(\varphi)$, instead of a
much more powerful Turing machine, we still cannot compute an appropriate
finite set of forbidden induced subgraphs for $\cls C$ from $\varphi$. On top
of it, Theorem~\ref{thm:igurgraph} implies that the size of forbidden
subgraphs for $\cls C$ cannot be bounded by any computable function in terms
of the size of $\varphi$.

There is an important precursor for Theorem~\ref{thm:igurgraph},
\begin{theo}[Gurevich's Theorem~\cite{gur84}]\label{thm:igur}
Let $f:\mathbb N\to \mathbb N$ be computable. Then there is an
\FO-sentence~$\varphi$ such that the class $\Mod(\varphi)$ of models of
$\varphi$ is closed under induced substructures but for every universal
sentence $\mu$ with $\Mod_{\fin}(\mu)= \Mod_{\fin}(\varphi)$ we have
$|\mu|\ge f(|\varphi|)$.
\end{theo}
\noindent Hence, Theorem~\ref{thm:igurgraph} can be viewed as the
graph-theoretic version of Theorem~\ref{thm:igur}.

Besides its importance in graph theory, Theorem~\ref{thm:igurgraph} is also
relevant in the context of algorithmic model theory. For algorithmic
applications, the \LT\ theorem provides a normal form (i.e., a universal
sentence) for any \FO-sentence preserved under induced substructures.
In~\cite{dawgrokresch07}, it is shown that on \emph{labelled trees} there is
no \emph{elementary bound} on the length of the equivalent universal sentence
in terms of the original one. We should point out that
Theorem~\ref{thm:igurgraph} is not comparable to Theorem~6.1
in~\cite{dawgrokresch07}, since our lower bound is uncomputable (and thus,
much higher than non-elementary) while the classes of graphs we construct in
the proof are dense (thus very far from trees).


\paragraph{Our technical contributions.}
For every vocabulary it is well known that the class of structures of this
vocabulary is \FO-interpretable in graphs (see for example~\cite{ebbflu99}).
So one might expect that Theorem~\ref{thm:Taitgraphs} and
Theorem~\ref{thm:igurgraph} can be derived easily from Tait's Theorem and
Gurevich's Theorem using the standard \FO-interpretations. However, an easy
analysis shows that those interpretations result in classes of graphs that
are not closed under induced subgraphs. So we introduce the notion of
\emph{strongly existential interpretation} which translates any class of
structures preserved under induced substructures to a class of graphs closed
under induced subgraphs. A lot of care is needed to construct strongly
existential interpretations.


\paragraph{Related research.}
Let us briefly mention some further results related to the \LT\ Theorem.
Essentially one could divide them into three categories: (a) The
\emph{positive results} showing that for certain classes $\cls C$ of finite
structures the analogue of the \LT\ Theorem holds if we restrict to
structures in $\cls C$. For example, this is the case if $\cls C$ is the
class of all finite structures of tree-width $k$ or less for some $k\in
\mathbb N$~\cite{atsdawgro08} or if $\cls C$ is the class of all finite
structures whose hypergraph satisfies certain properties~\cite{dur10}. (b)
Both just mentioned papers contain also \emph{negative results}, i.e, classes
for which the analogue of the \LT\ Theorem fails: For example,
in~\cite{atsdawgro08} this is shown for the class of finite planar graphs.
(c) The third category contains generalizations of the \LT\ Theorem. For
example, in~\cite{sanadscha16} the authors, for $k\ge 1$ consider sentences
of the form $\exists x_1\ldots \exists x_k\mu$, where $\mu$ is universal.
Then the role of the closure under induced substructures is taken over by a
semantic ``core property PS($k$)'' which for $k= 0$ coincides with closure
under induced substructures. Finally, we mention that in~\cite{dawsan20} the
authors strengthen Tait's result by showing that for every $n\ge 1$ there are
first-order definable classes of finite structures closed under substructures
which are not definable with $n$ quantifier alternations.

\paragraph{Organization of this paper.}
In Section~\ref{sec:pre} we fix some notations and recall or derive some
results about universal sentences we need in this paper. For the reader's
convenience, in Section~\ref{sec:idea} we include a proof of Tait's result.
Moreover, we prove a technical result, i.e., Proposition~\ref{pro:agr}, which
is an important tool in Gurevich's Theorem. We introduce the concept of
strongly existential interpretation in Section~\ref{sec:int} and show that
the results of the preceding section remain true under such interpretations.
We present an appropriate strongly existential interpretation for graphs (in
Section~\ref{sec:graph}). Hence, we get the results of Section~\ref{sec:idea}
for graphs. In Section~\ref{sec:gur} we first derive Gurevich's Theorem and
apply our interpretations to get the results for graphs. Finally, in
Section~\ref{sec:und}, we prove that various problems related to our results
are undecidable.

\section{Preliminaries}\label{sec:pre}

We denote by $\mathbb N$ the set of natural numbers greater or equal to 0.
For $n\in \mathbb N$ let $[n]:= \{1, 2, \ldots, n\}$.

\paragraph{First-order logic \FO.}
A \emph{vocabulary} $\tau$ is a finite set of relation symbols. Each relation
symbol has an \emph{arity}. A \emph{structure}~$\str A$ of vocabulary $\tau$,
or \emph{$\tau$-structure}, consists of a (finite or infinite) nonempty
set~$A$, called the \emph{universe} of $\str A$ and of an interpretation
$R^{\str A}\subseteq A^r$ of each $r$-ary relation symbol $R\in \tau$. If
$\str A$ and $\str B$ are $\tau$-structures, then \emph{$\str A$ is a
substructure of $\str B$}, denoted by $\str A\subseteq \str B$, if
$A\subseteq B$ and $R^{\str A}\subseteq R^{\str B}$, and \emph{$\str A$ is an
induced substructure of \str B}, denoted by $\str A\subseteq_\ind \str B$, if
$A\subseteq B$ and $R^{\str A}= R^{\str B}\cap A^r$, where~$r$ is the arity
of $R$. If, in addition, $A\subsetneq B$, then $\str A$ is an \emph{proper}
induced substructure of $\str B$. By $\Str[\tau]$ \big($\Str_\fin[\tau]$
\big) we denote the class of all (of all finite) $\tau$-structures.

\emph{Formulas} $\varphi$ of \emph{first-order logic} \FO\ of vocabulary
$\tau$ are built up from \emph{atomic formulas} $x_1= x_2$ and $Rx_1 \ldots
x_r$ (where $R\in \tau$ is of arity $r$ and $x_1, x_2, \ldots, x_r$ are
variables) using the boolean connectives $\neg$, $\wedge$, and $\vee$ and the
universal $\forall$ and existential $\exists$ quantifiers. A relation symbol
$R$ \emph{is positive (negative) in $\varphi$} if all atomic subformulas
$R\ldots$ in $\varphi$ appear in the scope of an \emph{even (odd)} number of
negation symbols.
By the notation $\varphi(\bar x)$ with $\bar x= x_1, \ldots, x_e$ we indicate
that the variables free in $\varphi$ are among $x_1, \ldots, x_e$. If then
$\str A$ is a $\tau$-structure and $a_1, \ldots, a_e\in A$, then $\str
A\models \varphi(a_1, \ldots, a_e)$ means that $\varphi(\bar x)$ holds in
$\str A$ if $x_i$ is interpreted by $a_i$ for $i\in [k]$.

A \emph{sentence} is a formula without free variables. For a sentence
$\varphi$ we denote by $\Mod(\varphi)$ the class of models of $\varphi$ and
$\Mod_\fin(\varphi)$ is its subclass consisting of the finite models of
$\varphi$. Sentences $\varphi$ and $\psi$ are \emph{equivalent} if
$\Mod(\varphi)= \Mod(\psi)$ and \emph{finitely equivalent} if
$\Mod_\fin(\varphi)= \Mod_\fin(\psi)$.

\paragraph{Graphs.}
Let $\tau_E:= \{E\}$ with binary $E$. For \emph{all} $\tau_E$-structures we
use the notation $G= (V(G), E(G))$ common in graph theory. Here $V(G)$, the
universe of $G$, is the set of vertices, and $E(G)$, the interpretation of
the relation symbol $E$, is the set of edges. The $\tau_E$-structure $G =
(V(G), E(G))$ is a \emph{directed graph} if $E(G)$ does not contain
self-loops, i.e., $(v,v)\notin E(G)$ for any $v\in V(G)$. If moreover
$(u,v)\in E(G)$ implies $(v,u)\in E(G)$ for any pair $(u,v)$, then $G$ is an
(undirected) \emph{graph}. The graph $H= (V(H), E(H))$ is an \emph{induced
subgraph} of $G$ if
\begin{eqnarray*}
V(H)\subseteq V(G) & \text{ and } &
 E(H)= E(G)\cap \big(V(H)\times V(H)\big).
\end{eqnarray*}
We denote by $\Graph$ and $\Graph_\fin$ the class of all graphs and the class
of finite graphs, respectively. Furthermore, for an $\FO[\tau_E]$-sentence
$\varphi$ by $\Graph(\varphi)$\big(and $\Graph_\fin(\varphi)$\big) we denote
the class of graphs (and the class of finite graphs) that are models of
$\varphi$.

\paragraph{Universal sentences and forbidden induced substructures.}
An \FO-formula is \emph{universal} if it is built up from atomic and negated
atomic formulas by means of the connectives $\wedge$ and $\vee$ and the
universal quantifier $\forall$. Often we say that a formula, say, containing
the connective $\to$ is universal if by replacing $\varphi\to \psi$ by $\neg
\varphi\vee \psi$ (and ``simple manipulations'') we get an equivalent
universal sentence. Every universal sentence $\mu$ is equivalent to a
sentence of the form $\forall x_1\ldots \forall x_k\, \mu_0$ for some $k\in
\mathbb N$ and some quantifier-free $\mu_0$ and moreover the length $|\mu|$
of $ \mu$ is at most $|\varphi|$. If in the definition of universal formula
we replace the universal quantifier by the existential one we get the
definition of an \emph{existential formula}.

One easily verifies that the class of models of a universal sentence is
closed under induced substructures. As already mentioned in the Introduction
for classes of graphs, \L o\'s~\cite{los55} and Tarski~\cite{tar54} proved:
\begin{theo}[\LT\ Theorem]\label{thm:ltp}
Let $\tau$ be a vocabulary and $\varphi$ an $\FO[\tau]$-sentence. Then
$\Mod(\varphi)$ is closed under induced substructures if and only if
$\varphi$ is equivalent to a universal sentence.
\end{theo}

We fix a vocabulary $\tau$. Let $\cls F$ be a finite set of finite
$\tau$-structures and denote by $\Forb(\cls F)$ (and $\Forb_\fin(\cls F)$)
the class of structures (of finite structures) that do not contain an induced
substructure isomorphic to a structure in $\cls F$. Clearly for finite sets
$F$ and $F'$ of finite $\tau$-structures we have
\begin{equation}\label{eq:foranti}
\text{if $\cls F\subseteq \cls F'$,
 then $\Forb(\cls F')\subseteq \Forb(\cls F)$.}
\end{equation}
We say that a class $\cls C$ of $\tau $-structures (of finite
$\tau$-structures) is \emph{definable by a finite set of forbidden induced
substructures} if there is a finite set $\cls F$ of finite structures such
that $\cls C= \Forb(\cls F)$ \big($\cls C= \Forb_\fin(\cls F)$\big).

Recall that $\tau_E= \{E\}$ with binary $E$.
\begin{eqnarray}\label{eq:axgr}
\varphi_{\textup{DG}}:= \forall x\neg Exx
 & \text{ and } &
\varphi_\Graph:= \forall x\neg Exx\wedge \forall x\forall y(Exy\to Eyx)
\end{eqnarray}
axiomatize the classes of directed graphs and of graphs, respectively. Let
the $\tau_E$-structures $H_0=(V(H_0), E(H_0))$ and $H_1=(V(H_1), E({H_1}))$
be given by
\begin{eqnarray*}
V(H_0):= \{1\}, \ E(H_0):= \big\{(1, 1)\big\}
 & \text{ and } &
V(H_1):= \{1, 2 \}, \ E(H_0):= \big\{(1, 2)\big\}.
\end{eqnarray*}
Then $\Forb\big(\{H_0\}\big)$ and $\Forb\big(\{H_0, H_1\}\big)$ are the class
of directed graphs and the class of graphs, respectively, i.e.,
$\Mod(\varphi_{\textup{DG}})= \Forb\big(\{H_0\}\big)$ and
$\Mod(\varphi_{\Graph})= \Forb\big(\{H_0, H_1\}\big)$.

The following result generalizes this simple fact and establishes the
equivalence between axiomatizability by a universal sentence and definability
by a finite set of forbidden induced substructures. For an arbitrary
vocabulary $\tau$, an $\FO[\tau]$-sentence $\varphi$, and $k\ge
1$ let
\begin{equation}\label{eq:deffis}
\cls F_k(\varphi):=
 \big\{\str A\in\Str[\tau] \bigmid
  \str A\models \neg\varphi \text{\ and $A=[\ell]$ for some $\ell\in[k]$}\big\}.
\end{equation}
Thus, $\cls F_k(\varphi)$ is, up to isomorphism, the class of structures with
at most $k$ elements which fail to be a model of $\varphi$. Note that $\cls
F_1(\varphi_{\textup{DG}})= \{H_0\}$ and $\cls F_1(\varphi_{\Graph})= \{H_0,
H_1\}$. Clearly, for a $\tau$-sentence we have:
\begin{align}\notag
\text{if $\Mod(\varphi)$ is closed under indu} & \text{ced substructures}, \\\label{eq:subfor}
 & \text{then $\Mod(\varphi)\subseteq \Forb(\cls F_k(\varphi))$ for all $k\ge 1$}.
\end{align}

\begin{prop}\label{pro:unifbk}
For a class $\cls C$ of $\tau$-structures and $k\ge 1$ the statements~(i)
and~(ii) are equivalent.
\begin{enumerate}
\item[(i)] $\cls C= \Mod(\mu)$ for some universal sentence $\mu:= \forall
    x_1\ldots \forall x_k\, \mu_0$ with quantifier-free~$\mu_0$.

\item[(ii)] $\cls C= \Forb(\cls F)$ for some finite set $\cls F$ of
    structures, all of at most $k$ elements.
\end{enumerate}
If (i) holds for $\mu$, then $\cls C= \Forb(\cls F_k(\mu))$.
\end{prop}

\proof (i) $\Rightarrow$ (ii) Let $\cls C=\Mod(\mu)$ for $\mu$ as in (i).
Then $\Mod(\mu)$ is closed under induced substructures and hence, $\cls C\subseteq
\Forb\big(\cls F_k(\mu)\big)$ by~\eqref{eq:subfor}.

Now assume that $\str A\notin \cls C$. Then $\str A\models \neg \mu$ and
hence there are $a_1, \ldots, a_k\in A$ with $\str A\models \neg \mu_0(a_1,
\ldots, a_k)$. For $\str B:= [a_1, \ldots, a_k]^{\str A}$, the substructure
of $\str A$ induced by $a_1, \ldots, a_k$, we have $\str B\models \neg
\mu_0(a_1,\ldots, a_k)$ (as $\mu_0$ is quantifier-free) and thus, $\str
B\models \neg\mu$. Therefore, $\str B$ is isomorphic to a structure in $\cls
F_k(\mu)$ and therefore, $\str A\notin \Forb\big(\cls F_k(\mu)\big)$.

\medskip
\noindent (ii) $\Rightarrow$ (i) Let the $\tau$-structure $\str A$ have at
most $k$ elements and let $a_1, \ldots, a_k$ be an enumeration of the
elements of $A$ (possibly with repetitions). Let $\delta(\str A; a_1, \ldots,
a_k)$ be the conjunction of all literals (i.e., atomic or negated atomic
formulas) $\lambda(x_1, \ldots, x_k)$ such that $\str A\models \lambda(a_1,
\ldots, a_k)$. Then for every $\tau$-structure $\str B$ and $b_1, \ldots,
b_k\in B$ we have
\begin{align}\notag
\str B\models \delta(\str A; a_1, \ldots, a_k)(b_1, \ldots, b_k)
 \iff & \text{ the clauses $\pi(a_i)= b_i$ for $i\in[k]$} \\\label{eq:puniso}
  &  \text{ define an isomorphism from $\str A$ onto $[b_1,\ldots, b_k]^{\str B}$.}
\end{align}
Now assume~(ii), i.e., $\cls C= \Forb(\cls F)$ for some finite set $\cls F$
of structures, all of at most $k$ elements. If $\cls F$ is empty, then $\cls
C=\Mod(\forall x\, x=x)$. Otherwise for every $\str A\in \cls F$ we fix an
enumeration $a^{\str A}_1, \ldots, a^{\str A}_k$ of the elements of $A$. We
set
\begin{equation*}
\mu:= \forall x_1\ldots \forall x_k
 \bigwedge_{\str A\in \cls F}\neg \delta(\str A; a^{\str A}_1, \ldots, a^{\str A}_k).
\end{equation*}
Then $\Forb(\cls F)= \Mod(\mu)$. In fact, assume first that $\str B\notin
\Mod(\mu)$. Then there are $b_1, \ldots, b_k\in B$ and an $\str A\in \cls F$
such that $\str B\models \delta(\str A; a^{\str A}_1, \ldots, a^{\str
A}_k)(b_1, \ldots, b_k)$. By~\eqref{eq:puniso}, then $\str A$ is isomorphic
to the induced substructure $[b_1, \ldots, b_k]^{\str B}$ of $\str B$; hence,
$\str B\notin \Forb(\cls F)$.

Now assume $\str B\notin \Forb(\cls F)$. Then there is an $\str A\in \cls F$
and elements $b_1, \ldots, b_k\in B$ such that the clauses $\pi(a^{\str
A}_i)= b_i$ for $i\in [k]$ define an isomorphism from $\str A$ onto $[b_1,
\ldots, b_k]^{\str B}$. By~\eqref{eq:puniso}, then $\str B\models \delta(\str
A; a^{\str A}_1, \ldots, a^{\str A}_k)(b_1, \ldots, b_k)$. Therefore, $\str
B\models \neg \mu$, i.e., $ \str{ B}\notin\Mod(\mu)$. \proofend



\begin{cor}\label{cor:vlem}
Let $\varphi$ be a $\tau$-sentence and $k\ge 1$. Then
\begin{eqnarray*}
\Mod(\varphi)= \Forb\big(\cls F_k(\varphi)\big) & \iff &
 \text{ $\varphi$ is equivalent to a universal sentence} \\
 & & \quad \text{ of the form $\forall x_1\ldots \forall x_k\, \mu_0$
   with quantifier-free~$\mu_0$.}
\end{eqnarray*}
\end{cor}

\noindent
By~\eqref{eq:foranti} and~\eqref{eq:subfor} we get:
\begin{cor}\label{cor:unimon}
If $\Mod(\mu)= \Forb\big(\cls F_k(\mu)\big)$ for some universal $\mu$ and
some $k\in \mathbb N$, then $\Mod(\mu)= \Forb\big(\cls F_\ell(\mu)\big)$ for
all $\ell\ge k$.
\end{cor}

\medskip
\begin{cor}\label{cor:unileq}
It is decidable whether two universal sentences are equivalent.
\end{cor}

\proof Let $\mu$ and $\mu'$ be universal sentences. W.l.o.g.\ we may assume
that $\mu= \forall x_1\ldots \forall x_k\, \mu_0$ and $\mu'= \forall
x_1\ldots \forall x_\ell\, \mu'_0$ with $k\le \ell$. By
Corollary~\ref{cor:vlem} and Corollary~\ref{cor:unimon}, we have
\begin{eqnarray*}
\Mod(\mu)= \Forb\big(\cls F_\ell(\mu)\big)
 & \text{ and } &
\Mod(\mu')= \Forb\big(\cls F_\ell(\mu')\big).
\end{eqnarray*}
Thus $\mu$ and $\mu'$ are equivalent if and only if $\cls F_\ell(\mu)= \cls
F_\ell(\mu')$. The right hand side of this equivalence is clearly decidable.
\proofend

\noindent
The last equivalence of this corollary shows:
\begin{cor}\label{cor:unilfeq}
For universal sentences $\mu$ and $\mu'$ we have
\[
\text{$\mu$ and $\mu'$ are equivalent}
 \iff \text{$\mu$ and $\mu'$ are finitely equivalent.}
\]
\end{cor}

\noindent
The following consequence of Corollary~\ref{pro:unifbk} will be used in the
next section.
\begin{cor}\label{cor:tluni}
Let $m,k\in \mathbb N$ with $m> k$ and let $\psi_0$ and $\psi_1$ be
$\FO[\tau]$-sentences. Assume that $\str A$ is a finite model of
$\psi_0\wedge \psi_1$ with at least $m$ elements and all its proper induced
substructures with at most~$k$ elements are models of $\psi_0\wedge \neg
\psi_1$. Then $\psi_0\wedge \neg \psi_1$ is not finitely equivalent to a
universal sentence of the form $\mu:= \forall x_1\ldots \forall x_k\,\mu_0$
with quantifier-free $\mu_0$.
\end{cor}

\proof For a contradiction assume $\Mod_\fin(\psi_0\wedge \neg \psi_1)=
\Mod_\fin(\mu)$ for $\mu$ as above. As $\Mod(\mu)= \Forb\big(\cls
F_k(\mu)\big)$ by Proposition~\ref{pro:unifbk}, we get (applying the finitely
equivalence of $\psi_0\wedge \neg \psi_1$ and $\mu$ to obtain the last
equality)
\[
\Mod_\fin(\psi_0\wedge \neg \psi_1)
 = \Mod_\fin(\mu)
 = \Forb_\fin\big(\cls F_k(\mu)\big)
 = \Forb_\fin\big(\cls F_k(\psi_0\wedge \neg \psi_1)\big).
\]
However, by the assumptions the structure $\str A$ is contained in
$\Mod_\fin(\psi_0\wedge \neg \psi_1)$ but not in the class $\Forb_\fin(\cls
F_k(\psi_0 \wedge \neg \psi_1))$. \proofend

\begin{rem}\label{rem:ltsub}
Let $\cls C$ be a class of $\tau$-structures closed under induced
substructures. For an $\FO[\tau]$-sentence $\varphi$ we set $\Mod_{\cls
C}(\varphi):= \{\str A\in \cls C \mid \str A\models \varphi\}$. We say that
the \emph{\LT\ Theorem holds for~$\cls C$} if for every $\FO[\tau]$-sentence
$\varphi$ such that the class $\Mod_{\cls C}(\varphi)$ is closed under
induced substructures there is a universal sentence $\mu$ such that
\[
\Mod_{\cls C}(\varphi)=\Mod_{\cls C}(\mu).
\]
The following holds:
\begin{quote}
{\em Let $\cls C$ and $\cls C'$ be classes of $\tau$-structures closed
under induced substructures with $\cls C'\subseteq \cls C$. Furthermore
assume that there is a universal sentence $\mu_0$ such that $\cls
C'=\Mod_{\cls C}(\mu_0)$. If the analogue of the \LT\ Theorem holds for
$\cls C$, then it holds for $\cls C'$, too}
\end{quote}
In fact, for every $\FO[\tau]$-sentence $\varphi$ we have $\Mod_{\cls
C'}(\varphi)=\Mod_{\cls C}(\mu_0\wedge \varphi)$. Hence, if $\Mod_{\cls
C'}(\varphi)$ is closed under induced substructures, then by assumption there
is a universal $\mu$ such that $\Mod_{\cls C}(\mu_0\wedge \varphi)=
\Mod_{\cls C}(\mu)$. Therefore, $\Mod_{\cls C'}(\varphi)=\Mod_{\cls C}(\mu)=
\Mod_{\cls C'}(\mu)$.
\end{rem}

\section{Basic ideas underlying the classical results}\label{sec:idea}

This section contains a proof of Tait's Theorem telling us that the analogue
of the \LT-Theorem fails if we only consider finite structures. Afterwards we
refine the argument to derive a generalization, namely
Proposition~\ref{pro:agr}, which is a key result to get Gurevich's Theorem.

We consider the vocabulary $\tau_0:= \{<, U_\min, U_\max, S\}$, where $<$ and
$S$ (the successor relation) are binary relation symbols and $U_\min$ and
$U_\max$ are unary.

Let $\varphi_0$~\label{pag:var0} be the conjunction of the universal
sentences
\begin{itemize}
\item $\forall x \neg x< x$, \ \ $\forall x\forall y (x<y\vee x=y\vee
    y<x)$, \quad $\forall x\forall y\forall z ((x<y\wedge y<z)\to x<z)$, \
    i.e., ``$<$ is an ordering''

\item $\forall x\forall y\big((U_\min\, x\to (x=y\vee x<y)\big)$ \ i.e.,
    ``every element in $U_\min$ is a minimum w.r.t.\ $<$''

\item $\forall x\forall y\big((U_\max\, x\to (x=y\vee y<x)\big)$ \ i.e.,
    ``every element in $U_\max$ is a maximum w.r.t.\ $<$''

\item $\forall xy(Sxy\to x<y)$

\item $\forall x\forall y\forall z(x<y<z\to \neg Sxz)$.
\end{itemize}
Note that from the axioms it follows that there is at most one element in
$U_\min$, at most one in $U_\max$, and that $S$ is a subset of the successor
relation w.r.t.\ $<$. We call \emph{$\tau_0$-orderings} the models of
$\varphi_0$.

For $\tau_0$-structures $\str A$ and $\str B$ we write $\str B\subseteq_<
\str A$ and say that $\str B$ is a \emph{$<$-substructure} of $\str A$ if
$\str A$ is a substructure of $\str B$ with $<^{\str B}= <^{\str A}\cap\,
(B\times B)$.

We remark that the relation symbols $U_\min, \ U_\max$, and $S$ are
negative in $\varphi_0$. Therefore we have:
\begin{lem}\label{lem:orind}
Let $\str B\subseteq_< \str A$. If $\str A\models \varphi_0$, then $\str
B\models \varphi_0$.
\end{lem}

Let
\begin{equation}\label{eq:var1}
\varphi_1:=\exists x\, U_\min\, x
 \wedge \exists x U_\max\, x\wedge \forall x\forall y(x<y \to \exists z Sxz).
\end{equation}
We call models of $\varphi_0\wedge\varphi_1$ \emph{complete
$\tau_0$-orderings}. Clearly, for every $k\ge 1$ there is a unique, up to
isomorphism, complete $\tau_0$-ordering with exactly $k$ elements. The next
lemma shows that all its proper $<$-substructures are models of
$\varphi_0\wedge\neg\varphi_1$.

\begin{lem}\label{lem:infb}
Let $\str A$ and $\str B$ be $\tau_0$-structures. Assume that $\str A\models
\varphi_0$ and $\str B$ is a finite $<$-substructure of~$\str A$ that is a
model of $\varphi_1$. Then $\str B= \str A$ (in particular, $\str A\models
\varphi_1$).
\end{lem}

\proof By the previous lemma we know that $\str B\models \varphi_0$. Let $B:=
\{b_1, \ldots, b_n\}$. As $<^{\str B}$ is an ordering, we may assume that
\[
b_1<^{\str B} b_2<^{\str B}\ldots <^{\str B} b_{n-1}<^{\str B} b_n.
\]
As $\str B\models (\varphi_0\wedge \varphi_1)$, we have $U_\min^{\str B}
b_1$, $U_\max^{\str B} b_n$, and $S^{\str B} b_ib_{i+1}$ for $i\in [n-1]$. As
$\str B\subseteq \str A$, everywhere we can replace the upper index $^{\str
B}$ by $^{\str A}$.

We show $A= B$: Let $a\in A$. By $\str A\models \varphi_0$, we have
$b_1\le^{\str A} a\le^{\str A} b_n$. Let $i\in [n]$ be maximal with $b_i
\le^{\str A} a$. If $i= n$, then $b_n= a$. Otherwise $b_i\le^{\str A}
a<^{\str A} b_{i+1}$. As $S^{\str A} b_ib_{i+1}$, we see that $b_i= a$ (by
the last conjunct of $\varphi_0$). Now $\str A= \str B$ follows from $\str
A\models \varphi_0$. \proofend

\begin{cor}\label{cor:infb}
Every proper $<$-substructure of a finite model of $\varphi_0\wedge
\varphi_1$ is a model of $\varphi_0\wedge \neg\varphi_1$.
\end{cor}

The class of finite $\tau_0$-orderings that are not complete is closed under
$<$-substructures but not axiomatizable by a universal sentence:
\begin{theo}[Tait's Theorem]\label{thm:ltf}
The class $\Mod_\fin(\varphi_0\wedge \neg\varphi_1)$ is closed under
$<$-substructures (and hence, closed under induced substructures) but
$\varphi_0\wedge \neg\varphi_1$ is not finitely equivalent to a universal
sentence.
\end{theo}

\proof $\Mod_\fin(\varphi_0\wedge \neg\varphi_1)$ is closed under
$<$-substructures: If $\str A\models \varphi_0\wedge \neg\varphi_1$ and $\str
B$ is a finite $<$-substructure of $\str A$, then $\str B\models \varphi_0$
(by Lemma~\ref{lem:orind}). If $\str B\models \neg\varphi_1$, we are done. If
$\str B\models \varphi_1$, then $\str A\models \varphi_1$ by
Lemma~\ref{lem:infb}, which contradicts our assumption $\str A\models
\neg\varphi_1$.

Let $k\in \mathbb N$. It is clear that there is a finite model $\str A$ of
$\varphi_0\wedge \varphi_1$ with at least $k+1$ elements. By
Corollary~\ref{cor:infb} every proper induced substructure of $\str A$ is a
model of $\varphi_0\wedge \neg\varphi_1$. Therefore, by
Corollary~\ref{cor:tluni}, the sentence $\varphi_0\wedge \neg\varphi_1$ is
not finitely equivalent to a universal sentence of the form $\mu:= \forall
x_1\ldots \forall x_k\,\mu_0$ with quantifier-free $\mu_0$. As $k$ was
arbitrary, we get our claim. \proofend

\begin{rem}\label{rem:sigma1}
A slight generalization of the previous proof shows that
$\Mod_\fin(\varphi_0\wedge \neg\varphi_1)$ is not even axiomatizable by a
$\Pi_2$-sentence, i.e., by a sentence $\chi$ of the form $\forall x_1\ldots
\forall x_k\exists y_1\ldots \exists y_\ell\, \chi_0$ for some $k, \ell\ge 1$
and quantifier-free $\chi_0$. In fact, assume that $\Mod_\fin(\varphi_0\wedge
\neg\varphi_1)= \Mod_\fin(\chi)$. Again we choose a finite model $\str A$ of
$\varphi_0\wedge \varphi_1$ with at least $k+1$ elements. Then $\str
A\not\models \chi$. Hence there are $a_1,\ldots, a_k\in A$ with $\str
A\models \neg\exists y_1\ldots \exists y_\ell\, \chi_0(a_1, \ldots, a_k)$.
Then $\str B\models \neg \exists y_1 \ldots \exists y_\ell\,
\chi_0(a_1,\ldots, a_k)$, where $\str B:= [a_1, \ldots, a_k]^{\str A}$ is the
substructure of $\str A$ induced by $a_1, \ldots, a_k$. Hence, $\str
B\not\models \chi$ and therefore, $\str B\not\models \varphi_0\wedge
\neg\varphi_1$. But this contradicts Corollary~\ref{cor:infb} as $\str B$ is
a proper induced substructure of~$\str A$.

Note that $\varphi_0\wedge \neg\varphi_1$ is (equivalent to) a
$\Sigma_2$-sentence, i.e., equivalent to the negation of a
$\Pi_2$-sentence.
%
\end{rem}

\noindent
We turn to a refinement of the previous statement that will be helpful to get
Gurevich's Theorem.
\begin{defn}\label{def:pair}\begin{enumerate}
\item[(a)] Let $\tau$ be obtained from the vocabulary $\tau_0$ by adding
    finitely many relation symbols ``in pairs,'' the \emph{standard} $R$
    together with its \emph{complement} $R^{\comp}$ (intended as the
    complement of $R$). The symbols $R$ and $R^{\comp}$ have the same arity
    and for our purposes we can restrict ourselves to unary or binary
    relation symbols (even though all results can be generalized to arbitrary
    arities). We briefly say that \emph{$\tau$ is obtained from $\tau_0$
    by adding pairs}.


\item[(b)] Let $\tau$ be obtained from $\tau_0$ by adding pairs. We say
    that $\varphi_{0\tau}$ \emph{is a $\tau$-extension of $\varphi_0$}
    (where $\varphi_0$ is as above) if it is a universal sentence such that
    \begin{itemize}
    \item[(i)] the sentence $\varphi_0$ is a conjunct of
        $\varphi_{0\tau}$,

    \item[(ii)] the sentence $\bigwedge_{R \textup{ standard}}\forall \bar
        x(\neg R\bar x\vee \neg R^{\comp}\bar x)$ is a conjunct of
        $\varphi_{0\tau}$,

    \item[(iii)] besides $<$ all relation symbols are negative in
        $\varphi_{0\tau}$ (if this is not the case for some new $R$ or
        $R^{\comp}$, the idea is to replace any positive occurrence of $R$
        or $R^{\comp}$ by $\neg R^{\comp}$ and $\neg R$, respectively). For
        instance, we replace a subformula
        \begin{eqnarray*}
        x<y \wedge R xy \ \
         & \text{by} & \ \
        x<y \wedge \neg R^{\comp} xy.
        \end{eqnarray*}
\end{itemize}

\item[(c)] Let $\tau$ be obtained from $\tau_0$ by adding pairs. Then we
    set
    \begin{equation}\label{eq:v1tau}
    \varphi_{1\tau}:= \varphi_1\wedge\bigwedge_{R \textup{ standard}}
     \forall \bar x(R\bar x\vee R^{\comp}\bar x),
    \end{equation}
    where $\varphi_1$ is as above \big(see \eqref{eq:var1}\big).
\end{enumerate}
\end{defn}

\noindent For a $\tau$-structure $\str B$ with $\str B\models
\varphi_{0\tau}\wedge \varphi_{1\tau}$ we have
\[
\str B\models\bigwedge_{R \textup{ standard}} \Big(\forall \bar
        x(\neg R\bar x\vee \neg R^{\comp}\bar x) \wedge
     \forall \bar x(R\bar x\vee R^{\comp}\bar x)\Big).
\]
Hence,
\begin{equation}\label{eq:comp}
\text{if $\str B\models \varphi_{0\tau}\wedge \varphi_{1\tau}$,
then $(R^{\comp})^{\str B}$ is the complement of $R^{\str B}$ for standard $R\in\tau$}.
\end{equation}
Now we derive the analogues of Lemma~\ref{lem:orind}--Theorem~\ref{thm:ltf}
essentially by the same proofs.
\begin{lem}\label{lem:orindm}
Let $\tau$ be obtained from $\tau_0$ by adding pairs and let
$\varphi_{0\tau}$ be an extension of $\varphi_0$. If $\str B\subseteq_< \str
A$ and $\str A\models \varphi_{0\tau}$, then $\str B\models \varphi_{0\tau}$.
\end{lem}

\proof By Definition~\ref{def:pair}\,(b)\,(iii) all relation symbols distinct
from $<$ are negative in $\varphi_{0\tau}$. \proofend

\begin{lem}\label{lem:infm}
Let $\tau$ be obtained from $\tau_0$ by adding pairs and let
$\varphi_{0\tau}$ be an extension of $\varphi_0$. Assume that $\str A\models
\varphi_{0\tau}$ and that the finite $<$-substructure \str B of $\str A$ is a
model of $\varphi_{1\tau}$. Then $\str B= \str A$ \big(in particular, $\str
A\models \varphi_{1\tau}$\big).
\end{lem}

\proof Let $\str A\upharpoonright \tau_0$ (and $\str B\upharpoonright
\tau_0$) be the $\tau_0$-structure obtained from $\str A$ (from $\str B$) by
removing all relations in $\tau\setminus \tau_0$.

By Lemma~\ref{lem:infb} we know that $\str B\upharpoonright \tau_0= \str
A\upharpoonright \tau_0$. Furthermore, $\str B\models \varphi_{0\tau}$ by the
previous lemma; thus, $\str B\models \varphi_{0\tau}\wedge \varphi_{1\tau}$.
Hence, by~\eqref{eq:comp}, $(R^{\comp})^{\str B}$ is the complement of
$R^{\str B}$ for standard $R$. Clearly, $R^{\str B}\subseteq R^{\str A}$ and
$(R^{\comp})^{\str B}\subseteq (R^{\comp})^{\str A}$. As $A= B$ and $\str A$
is a model of the sentence $\bigwedge_{R \textup{ standard}}\forall \bar
x(\neg R\bar x\vee \neg R^{\comp}\bar x)$, we get $R^{\str B}= R^{\str A}$
and $(R^{\comp})^{\str B}= (R^{\comp})^{\str A}$. \proofend

\begin{cor}\label{cor:infm}
Every proper $<$-substructure of a finite model of $\varphi_{0\tau}\wedge
\varphi_{1\tau}$ is a model of $\varphi_{0\tau}\wedge \neg\varphi_{1\tau}$.
\end{cor}

By replacing in the proof of Tait's Theorem the use of Lemma~\ref{lem:orind},
Lemma~\ref{lem:infb}, and Corollary~\ref{cor:infb} by Lemma~\ref{lem:orindm},
Lemma~\ref{lem:infm}, and Corollary~\ref{cor:infm} respectively, we get:
\begin{lem}\label{lem:exttai}
Let $\tau$ be obtained from $\tau_0$ by adding pairs and let
$\varphi_{0\tau}$ be an extension of $\varphi_0$. The class
$\Mod_\fin(\varphi_{0\tau} \wedge \neg\varphi_{1\tau})$ is closed under
$<$-substructures (and hence, closed under induced substructures) but
$\varphi_{0\tau}\wedge \neg\varphi_{1\tau}$ is not finitely equivalent to a
universal sentence.
\end{lem}

Perhaps the reader will ask why we do not introduce for $<$ the ``complement
relation symbol''~$<^\comp$ and add the corresponding conjuncts to
$\varphi_{0\tau}$ and $\varphi_{1\tau}$ (or, to $\varphi_0$ and $\varphi_1$)
in order to get a result of the type of Lemma~\ref{lem:infm} (or already of
the type of Lemma~\ref{lem:infb}) where we can replace ``$ <$-substructure''
by ``substructure.'' The reader will realize that corresponding proofs of $B=
A$ break down.

\medskip
The next proposition provides a uniform way to construct \FO-sentences that
are only equivalent to universal sentences of large size, which is the core
of the proof of Gurevich's Theorem.

\begin{prop}\label{pro:agr}
Again let $\tau$ be obtained from $\tau_0$ by adding pairs and
$\varphi_{0\tau}$ be an extension of $\varphi_0$. Let $m\ge 1$ and $\gamma$
be an $\FO[\tau]$-sentence such that
\begin{equation}\label{eq:agr}
\text{$\varphi_{0\tau}\wedge \varphi_{1\tau}\wedge \gamma$ has
no infinite model but a finite model with at least $m$ elements}.
\end{equation}
For
\[
\chi:= \varphi_{0\tau}\wedge (\varphi_{1\tau}\to \neg \gamma)
\]
the statements (a) and (b) hold.
\begin{itemize}
\item[(a)] The class $\Mod(\chi)$ is closed under $<$-substructures.

\item[(b)] If $\mu:=\forall x_1\ldots \forall x_k\,\mu_0$ with
    quantifier-free $\mu_0$ is finitely equivalent to $\chi$, then $k\ge
    m$.
\end{itemize}
\end{prop}

\proof (a) Let $\str A\models \chi$ and $\str B\subseteq_< \str A$. Thus,
$\str B\models \varphi_{0\tau}$. If $\str B\not \models \varphi_{1\tau}$, we
are done. Assume $\str B \models \varphi_{1\tau}$. In case $B$ is infinite,
we conclude by~\eqref{eq:agr} that $\str B$ is a model of $\neg \gamma$ and
hence of $\chi$. Otherwise $B$ is finite; then $\str B= \str A$ (by
Lemma~\ref{lem:infm}) and thus, $\str B\models \chi$.
%

\medskip
\noindent (b) According to~\eqref{eq:agr} there is a finite model $\str A$ of
$\varphi_{0\tau}\wedge \varphi_{1\tau}\wedge \gamma$, i.e., of
$\varphi_{0\tau}\wedge \neg (\varphi_{1\tau}\to \neg\gamma)$, with at least
$m$ elements. By Corollary~\ref{cor:infm} every proper induced substructure
of \str A is not a model of $\varphi_{1\tau}$ and therefore, it is a model of
$\varphi_{0\tau}\wedge (\varphi_{1\tau}\to \neg\gamma)$. Hence by
Corollary~\ref{cor:tluni}, $\varphi_{0\tau}\wedge (\varphi_{1\tau}\to
\neg\gamma)$ is not finitely equivalent to a universal sentence of the form
$\mu:=\forall x_1\ldots \forall x_k\,\mu_0$ with $k< m$ and quantifier-free $
\mu_0$. \proofend

\begin{rem}\label{rem:sigma1m}
We can strengthen the statement~(b) of the preceding proposition to:
\begin{quote}
{\em If the $\Pi_2$-sentence $\forall x_1\ldots \forall x_k\exists
y_1\ldots \exists y_\ell\; \chi_0$ with quantifier-free $\chi_0$ is
finitely equivalent to $\chi$, then $k\ge m$.}
\end{quote}
The proof is similar to that of the result in Remark~\ref{rem:sigma1} and is
left to the reader.
\end{rem}

\section{The general machinery: strongly existential interpretations}\label{sec:int}

We show that appropriate interpretations preserve the validity of Tait's
theorem and of the statement of Proposition~\ref{pro:agr}. Later on these
interpretations will allow us to get versions of the results for graphs.

\medskip
Let $\tau_E:= \{E\}$ with binary $E$. As already remarked in the
Preliminaries for all $\tau_E$-structures we use the notation $G= (V(G),
E(G))$ common in graph theory.

Let $\tau$ be obtained from $\tau_0$ by adding pairs. Furthermore, let $I$ be
an interpretation of \emph{width} $2$ (we only need this case) of
$\tau$-structures in $\tau_E$-structures. This means that~$I$ assigns to
every unary relation symbol $T\in \tau$ an $\FO[\tau_E]$-formula
$\varphi_T(x_1, x_2)$ and to every binary relation symbol $T\in \tau$ an
$\FO[\tau_E]$-formula $\varphi_T(x_1, x_2, y_1, y_2)$; moreover, $I$ selects
an $\FO[\tau_E]$-formula $\varphi_\univ(x_1, x_2)$.

Then $I$ assigns to every $\tau_E$-structure $G$ with $G\models \exists \bar
x\varphi_\univ(\bar x)$ a $\tau$-structure $G_I$,
which we often denote by $\str O_I(G)$, defined by
\begin{itemize}
\item $O_I(G):= \big\{\bar a\in V(G)\times V(G) \bigmid G\models
    \varphi_\univ(\bar a) \big\}$

\item $T^{O_I(G)}:= \big\{\bar a\in O_I(G) \bigmid G \models \varphi_T(\bar a)
    \big\}$ \ for unary $T\in \tau$

\item $T^{O_I(G)}:= \big\{(\bar a,\bar b)\in O_I(G)\times O_I(G) \bigmid G
    \models \varphi_T(\bar a, \bar b)\big\}$ \ for binary $T\in \tau$.
\end{itemize}
As the interpretation $I$ is of width $2$, we have
\begin{equation}\label{eq:sgoq}
|O_I(G)|\le |V(G)|^2.
\end{equation}
Recall that for every sentence $\varphi\in \FO[\tau]$ there is a sentence
$\varphi^I \in\FO[\tau_E]$ such that for all $\tau_E$-structures $G$ with
$G\models \exists \bar x\varphi_\univ(\bar x)$ we have
\begin{equation}\label{eq:obun}
\left(G_I=\right)\; \str O_I(G)\models \varphi \iff G\models \varphi^I.
\end{equation}
For example,
for the sentence $\varphi= \forall x\forall y\, Txy$ we have
\[
\varphi^I= \forall \bar x \Big(\varphi_{\univ}(\bar x)
\to \forall \bar y \big(\varphi_{\univ}(\bar y) \to \varphi_T(\bar x, \bar y)\big)\Big).
\]
Furthermore there is a constant $c_I\in\mathbb N$ such that for all
$\varphi\in \FO[\tau]$,
\begin{equation}\label{eq:cci}
|\varphi^I|\le c_I\cdot |\varphi|.
\end{equation}

\begin{defn}\label{def:strex}
Let $\tau$ be obtained from $\tau_0$ by adding pairs and let $I$ be an
interpretation of $\tau_0$-structures in $\tau_E$ as just described. We say
that $I$ is \emph{strongly existential} if all formulas of $I$ are
existential and $\varphi_<$ is even quantifier-free.
\end{defn}

\begin{lem}\label{lem:phi0}
Let $\tau$ be obtained from $\tau_0$ by adding pairs and let
$\varphi_{0\tau}$ be an extension of $\varphi_0$. Then for every strongly
existential interpretation $I$ the sentence $\varphi^I_{0\tau}$ is
(equivalent to) a universal sentence.
\end{lem}

\proof The claim holds as all relation symbols distinct from $<$ are
negative in $\varphi_{0\tau}$. For example, for $\varphi:=\forall x\forall
y \big(U_\min\, x\to (x=y\vee x<y)\big)$, we have
\[
\varphi^I=\forall \bar x
 \Big(\varphi_\univ(\bar x)\to
  \forall \bar y \big(\varphi_\univ(\bar y)
   \to (\varphi_{U_\min}(\bar x)\to ((x_1= y_1\wedge x_2= y_2) \vee \varphi_<(\bar x,\bar y)))\big)\Big).
\benda
\]

\noindent The following result shows that strongly existential
interpretations preserve induced substructures in such a way that we can
translate the results of the preceding section to the actual context.

\begin{lem}\label{lem:unim}
Assume that $I$ is strongly existential. Then for all $\tau_E$-structures $G$
and $H$ with $H\subseteq_\ind G$ and $O_I(H)\ne \emptyset$, we have $\str O_I(H)
\subseteq_< \str O_I(G)$.
\end{lem}

\proof As $\varphi_\univ$ is existential, we have $O_I(H)\subseteq O_I(G)$.
Let $T\in \tau$ be distinct from $<$ and $\bar b\in T^{\str O_I(H)}$. Then
$H\models \varphi_T(\bar b)$. As $\varphi_T$ is existential, $G\models
\varphi_T(\bar b)$ and thus, $\bar b\in T^{\str O_I(G)}$. Moreover, for $\bar
b, \bar b'\in O_I(H)$ we have
\begin{eqnarray*}
\bar b<^{\str O_I(H)}\bar b'
 & \iff & H\models \varphi_<(\bar b,\bar b') \\
 & \iff & G\models \varphi_<(\bar b,\bar b')
  \qquad \text{(as $H\subseteq_\ind G$ and $\varphi_<$ is quantifier-free)} \\
 & \iff & \bar b<^{\str O_I(G)}\bar b'.
\end{eqnarray*}
Putting all together we see that $\str O_I(H)\subseteq_< \str O_I(G)$.
\proofend

We obtain from Lemma~\ref{lem:infm} the corresponding result in our
framework.
\begin{lem}\label{lem:infgm}
Assume that $I$ is strongly existential. Let $\varphi_{0\tau}$ be an
extension of $\varphi_0$. Let $G$ be a $\tau_E$-structure and $G\models
\varphi_{0\tau}^I$. Let $H\subseteq_\ind G$ with finite $O_I(H)$. If $H\models
\varphi_{1\tau}^I$, then $\str O_I(H)= \str O_I(G)$ and $G\models
\varphi^I_{1\tau}$.
\end{lem}

\proof As $H\models \varphi_{1\tau}^I$, in particular $H\models (\exists x\,
U_\min\, x)^I$; thus, $O_I(H)\ne \emptyset$. Therefore, $\str
O_I(H)\subseteq_< \str O_I(G)$ by Lemma~\ref{lem:unim}. By assumption
and~\eqref{eq:obun}, $\str O_I(G)\models \varphi_{0\tau}$ and $\str
O_I(H)\models \varphi_{1\tau}$. As $O_I(H)$ is finite, Lemma~\ref{lem:infm}
implies $\str O_I(H)= \str O_I(G)$, and in particular $\str O_I(G)\models
\varphi_{1\tau}$. Hence, $G\models \varphi^I_{1\tau}$ by~\eqref{eq:obun}.
\proofend

We now prove for strongly existential interpretations two results,
Proposition~\ref{pro:ltfm} corresponds to Tait's Theorem
(Theorem~\ref{thm:ltf}), and Proposition~\ref{pro:grfm} corresponds to
Proposition~\ref{pro:agr} (relevant to Gurevich's Theorem). In our
application of these results to graphs in the next section the sentence
$\psi$ will be $\forall x\neg Exx\wedge \forall x\forall y (Exy\to Eyx)$,
i.e., the sentence $\varphi_\Graph$ \big(cf.~\eqref{eq:axgr}\big)
axiomatizing the class of graphs.

\begin{prop}\label{pro:ltfm}
Let $\psi$ be a universal $\tau_E$-sentence. Assume that the interpretation
$I$ of $\tau_0$-structures in $\tau_E$-structures is strongly existential.
Furthermore, assume that for every sufficiently large finite complete
$\tau_0$-ordering $\str A$ there is a finite $\tau_E$-structure~$G$ with
$\str O_I(G)\cong \str A$ and $G\models \psi$. Then there is an
$\FO[\tau_E]$-sentence $\varphi$ such that $\Mod_\fin(\psi\wedge \varphi)$ is
closed under induced substructures, but $\psi\wedge \varphi$ is not finitely
equivalent to a universal sentence.

As $\varphi$ we an take the sentence
\[
\varphi:= \forall \bar x\neg \varphi_\univ(\bar x)
 \vee \big(\varphi^I_0\wedge \neg \varphi^I_1\big)
\]
\big(for the definition of $\varphi_0$ and $\varphi_1$ see
page~\pageref{pag:var0} and~\eqref{eq:var1}, respectively\big).
\end{prop}

\proof First we verify that the class $\Mod_\fin(\psi\wedge \varphi)$ is
closed under induced substructures. Assume $G\models \psi\wedge \varphi$ and
$H\subseteq_\ind G$. Since~$\psi$ is universal, we have $H\models\psi$. If
$G\models \forall \bar x\neg \varphi_\univ(\bar x)$, then $H\models \forall
\bar x\neg \varphi_\univ(\bar x)$. Now assume that $G\models
\varphi^I_0\wedge \neg\varphi^I_1$. Then $H\models \varphi^I_0$, as
$\varphi_0^I$ is universal by Lemma~\ref{lem:phi0}. If $H\models \forall \bar
x\neg \varphi_\univ(\bar x)$ or $H\models \neg\varphi^I_1$, we are done.
Otherwise $O_I(H)\ne\emptyset$ and $H\models \varphi^I_1$. Then $G\models
\varphi^I_1$ (see Lemma~\ref{lem:infgm}), a contradiction.
\medskip

\noindent Finally we show that for every $k\in \mathbb N$ the sentence
$\psi\wedge \varphi$ is not finitely equivalent to a sentence of the form
$\mu= \forall z_1\ldots \forall z_k\,\mu_0$ with quantifier-free $\mu_0$. Let
\[
\str A:= \big(A, <^{\str A}, U_\min^{\str A}, U_\max^{\str A}, S^{\str A}\big)
\]
be a complete $ \tau_0$-ordering with at least $k^2+1$ elements. In
particular, $\str A\models \varphi_0\wedge \varphi_1$. By assumption we can
choose $\str A$ in such a way that there is a finite $\tau_E$-structure $G$
such that $\str O_I(G)\cong \str A$ and $G\models \psi$. Then $\str
O_I(G)\models \varphi_0\wedge \varphi_1$, hence, $G\models \varphi_0^I\wedge
\varphi^I_1$. Thus $G\models \psi \wedge \neg\varphi$. As $|O_I(G)|= |A|\ge
k^2+ 1$, the graph $G$ must contain more than $k$ elements
by~\eqref{eq:sgoq}.

We want to show that every induced substructure of $G$ with at most $k$
elements is a model of $\psi\wedge \varphi$. Then the result follows from
Corollary~\ref{cor:tluni}. So let $H$ be an induced substructure of $G$ with
at most~$k$ elements. Clearly, $H\models (\psi\wedge \varphi^I_{0})$. If
$H\models \forall \bar x\neg \varphi_\univ(\bar x)$ or $H\models \neg
\varphi^I_{1}$, we are done. Otherwise $O_I(H)\ne \emptyset$ and $H\models
\varphi^I_{1}$. Then, Lemma~\ref{lem:infgm} implies $O_I(H)=O_I(G)$. Recall
$|V(H)|\le k$, so $O_I(H)$ has at most~$k^2$ elements by~\eqref{eq:sgoq}, a
contradiction as $|O_I(G)|\ge k^2+1$. \proofend
%

\begin{prop}\label{pro:grfm}
Assume that $\psi$ is a universal $\tau_E$-sentence. Let $\tau$ be obtained
from $\tau_0$ by adding pairs and let $\varphi_{0\tau}$ be an extension of
$\varphi_0$. Let $I$ be a strongly existential interpretation of
$\tau$-structures in $\tau_E$-structures with the property that for every
finite $\tau$-structure $\str A$, which is a model of $\varphi_{0\tau}\wedge
\varphi_{1\tau}$, there is a finite $\tau_E$-structure $G$ with $\str
O_I(G)\cong \str A$ and $G\models \psi$.

Let $m\ge 1$ and $\gamma$ be an $\FO[\tau]$-sentence such that
\begin{equation}\label{eq:agrm}
\text{$\varphi_{0\tau}\wedge \varphi_{1\tau}\wedge \gamma$
 has no infinite model but a finite model with at least $m$ elements}.
\end{equation}
For
\begin{equation}\label{eq:agrvar}
\rho:=
 \forall \bar x\neg \varphi_\univ(\bar x)
  \vee \big(\varphi_{0\tau}\wedge (\varphi_{1\tau}\to \neg \gamma)\big)^I
\end{equation}
the statements (a) and (b) hold.
\begin{itemize}
\item[(a)] The class $\Mod(\psi\wedge\rho)$ is closed under induced
    substructures.

\item[(b)] If $\mu:= \forall x_1\ldots \forall x_k\,\mu_0$ with
    quantifier-free $\mu_0$ is finitely equivalent to $\psi\wedge \rho$,
    then $k^2\ge m$.
\end{itemize}
\end{prop}

\proof (a) Assume that $G\models \psi\wedge\rho$ and $H\subseteq_\ind G$.
Clearly $H\models \psi$. If $H\models \forall \bar x\neg \varphi_\univ(\bar
x)$, then we are done. Otherwise, the universe of $\str O_I(H)$ and hence,
that of $\str O_I(G)$, are not empty. Then $G\models \varphi^I_{0\tau}$ and
as $H\subseteq_\ind G$, we have $H\models \varphi^I_{0\tau}$ by
Lemma~\ref{lem:phi0}.

If $H\not\models \varphi^I_{1\tau}$, we are done. Otherwise, $H\models
\varphi^I_{1\tau}$. If $H_I$ is infinite, then $H_I\models \neg\gamma$
by~\eqref{eq:agrm} and we are again done. If $H_I$ is finite, then $\str
O_I(H)= \str O_I(G)$ by Lemma~\ref{lem:infgm}. Thus $\str O_I(G)\models
\varphi_{1\tau}$ and hence, $\str O_I(G)\models \neg \gamma$ as $G\models
\rho$. Therefore, $\str O_I(H)\models \neg \gamma$ and thus, $H\models \rho$.

\medskip
\noindent (b) By~\eqref{eq:agrm} there is a finite model $\str A$ of
$\varphi_{0\tau}\wedge \varphi_{1\tau}\wedge \gamma$ with at least $m$
elements. By assumption there is a finite $\tau_E$-structure~$G$ with $\str
O_I(G)\cong \str A$ and $G\models \psi$. Clearly, $G\models \neg \forall \bar
x\neg \varphi_\univ(\bar x)$ and $G\models (\varphi_{0\tau}\wedge
\varphi_{1\tau}\wedge \gamma)^I$. Hence, $G\models \psi\wedge\neg \rho$.
Assume that $k^2< m$. We want to show that every induced substructure of $G$
with at most $k$ elements is a model of $\psi\wedge \rho$. Then the claim~(b)
follows from Corollary~\ref{cor:tluni}.

So let $H$ be an induced substructure of $G$ with at most $k$ elements.
Clearly, $H\models (\psi\wedge \varphi^I_{0\tau})$. If $H\models \forall \bar
x\neg \varphi_\univ(\bar x)$ or $H\models \neg \varphi^I_{1\tau}$, we are
done. Otherwise $O_I(H)\ne \emptyset$ and $H\models \varphi^I_{1\tau}$. Then,
$O_I(H)= O_I(G)$ by Lemma~\ref{lem:infgm}. This leads to a contradiction, as
$O_I(H)$ has at most $k^2$ elements by~\eqref{eq:obun}, while $O_I(G)$ has
$m$ elements and we assumed $k^2< m$. \proofend

\begin{rem}\label{rem:sigma1n}
The results corresponding to Remark~\ref{rem:sigma1} and
Remark~\ref{rem:sigma1m} are valid for Proposition~\ref{pro:ltfm} and
Proposition~\ref{pro:grfm} too. In particular, the sentence
$\psi\wedge\varphi\ \big(=\psi\wedge \forall \bar x\neg \varphi_\univ(\bar x)
\vee \big(\varphi^I_0\wedge \neg \varphi^I_1\big)\big)$ is not equivalent to
a $\Pi_2$-sentence. Furthermore $\psi\wedge \varphi$ itself is equivalent to
a $\Sigma_2$-sentence. In fact, as all relation symbols besides $<$ are
negative in $\varphi_0$, the sentence $\varphi_0^I$ is universal. Moreover,
as $U_\min$, $U_\max$, and $S$ are positive in $\varphi_1$, the sentence
$\varphi_1^I$ (as $\varphi_1$) is equivalent to a $\Pi_2$-sentence. Hence
$\psi\wedge\varphi$ is equivalent to a $\Sigma_2$-sentence.
\end{rem}

\section{Tait's Theorem for finite graphs}\label{sec:graph}

In this section we introduce a strongly existential interpretation, which
allows us to get Tait's Theorem for graphs. The corresponding result for
Gurevich's Theorem will be derived in~Section~\ref{sec:gur}.

\begin{figure}
\centering
\input{pathear.tex}
\caption{A path of length 6 with a $4$-ear.}
\label{fig:pathear}
\end{figure}
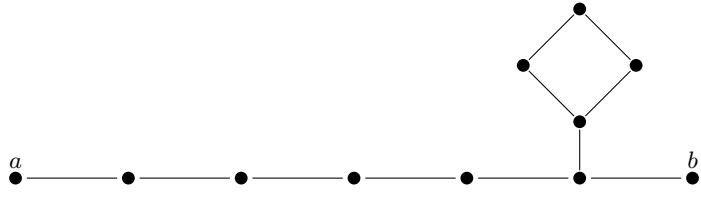

\medskip
We first introduce a further concept. Let $G$ be a graph and $a,b\in V(G)$.
For $r,s\ge 3$ \emph{a path from vertex $a$ to vertex $b$ of length $r$ with
an $s$-ear} is a path between $a$ and $b$ with a cycle of length $s$; one
vertex of this cycle is adjacent to the vertex adjacent to $b$ on the path.
Figure~\ref{fig:pathear} is a path from $a$ to $b$ of length $6$ with a
$4$-ear.

\begin{lem}\label{lem:path-cycle}
For $r,s\ge 3$ there are quantifier-free formulas $\varphi_{cr}(x, \bar z)$
and $\varphi_{pe,r,s}(x, y, \bar z, \bar w)$ such that for all graphs $G$ we
have
\begin{itemize}
\item[(a)] $G\models \varphi_{cr}(a, \bar u)\iff \bar u$ is a cycle of
    length $r$ containing $a$.

\item[(b)] $G\models \varphi_{pe,r,s}(a,b,\bar u,\bar v)\iff \bar u$ is
    path from $a$ to $b$ of length $r$ with the $s$-ear $\bar v$.
\end{itemize}
\end{lem}

\proof (a) We can take as $\varphi_{cr}(x, z_1,\ldots, z_r)$ the formula
\begin{equation*}
x=z_1 \wedge Ez_rz_1
  \wedge \bigwedge_{1\le i<r} Ez_iz_{i+1}
   \wedge \bigwedge_{1\le i<j\le r}\neg z_i= z_j.
\end{equation*}

\noindent (b) We can take as $\varphi_{pe,r,s}(x, y, z_0,\ldots, z_r, w_1,
\ldots, w_s)$ the formula
\begin{align*}
x= z_0 \wedge y=z_r\wedge \bigwedge_{0\le i<r-1}Ez_iz_{i+1}
 \wedge \bigwedge_{0\le i< j\le r}\neg z_i&= z_j
  \wedge \bigwedge_{0\le i\le r,\ j\in[s]}\neg z_i=w_j \\
  & \wedge \varphi_{cs}(w_1,w_1,\ldots, w_r)
   \wedge Ez_{r-1}w_1.
    \benda
\end{align*}
To understand better how we obtain the desired interpretation we first assign
to every complete $\tau_0$-ordering $\str A$, i.e., to every model of
$\varphi_0\wedge \varphi_1$, a $\tau_E$-structure $G:= G(\str A)$ which is a
graph.

In a first step we extend $\str A$ to a $\tau^*_0$-structure $\str A^*$,
where $\tau^*_0:= \tau_0\cup \{B,C, L, F\}$ in the following way. Here $B,C$
are unary and $L,F$ are binary relation symbols.

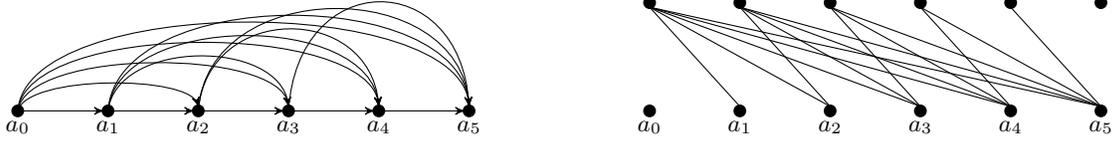
\begin{figure}
\centering
\input{relationF.tex}
\caption{Turning an ordering to the relation $F$.}
\label{fig:relationF}
\end{figure}

For every \emph{original} (or, \emph{basic}) element $a$, i.e., for every
$a\in A$, we introduce a new element $a'$, the \emph{companion of $a$}. We
set
\begin{itemize}
\item $A^*:= A\cup \{a' \mid a\in A\}$,

\item $B^{\str A^*}:= A$, \qquad $C^{\str A^*}:= \{a' \mid a\in A\}$,

\item $L^{\str A^*}:= \big\{(a,a')\bigmid a\in A\big\}, \qquad F^{\str
    A^*}:= \big\{(a',b),(b,a') \bigmid a,b\in A,\ a<^{\str A} b\big\}$.
\end{itemize}
Note that the relation $F$ is irreflexive and symmetric, i.e.,
$\big(A^*,F^{\str A^*}\big)$ is already a graph, which is illustrated by
Figure~\ref{fig:relationF}.
Observe that $F$ contains the whole information of the ordering $<^{\str A}$
up to isomorphism.

We use $\str A^*$ to define the desired graph $G= G(\str A)$. The vertex set
$V(G)$ contains the elements of~$A^*$, and the edge relation $E(G)$ contains
$F^{\str A^*}$. Furthermore $G$ contains just all the vertices and edges
required by the following items:
\begin{itemize}
\item To $a\in U_\min^{\str A}$ we add a cycle of length $5$ consisting of
    new vertices, i.e., not in $A^*$ (besides $a$).

\item To $a\in U_\max^{\str A}$ we add a cycle of length $7$ consisting of
    new vertices (besides $a$).

\item To $a\in B^{\str A^*}$ we add a cycle of length $9$ consisting of new
    vertices (besides $a$).

\item To $a\in C^{\str A^*}$ we add a cycle of length $11$ consisting of
    new vertices (besides $a$).

\item To $(a,b)\in S^{\str A}$ we add a path from $a$ to $b$ of length $17$
    with a $13$-ear consisting of new vertices (besides $a$ and~$b$).

\item To $(a,a')\in L^{\str A^*}$ we add a path from $a$ to $a'$ of length
    $17$ with a $15$-ear consisting of new vertices (besides $a$ and~$a'$).
\end{itemize}
Hereby we meant by ``add a cycle'' or ``add a path with an ear'' that we only
add the edges required by the corresponding formulas
in~Lemma~\ref{lem:path-cycle}.
\medskip

\noindent To ease the discussion, we divide cycles in $G\ (= G(\str A))$ into
four categories.
%

\medskip
\noindent [{\em $F$-cycle}] \ These are cycles in $\big(A^*,F^{\str
A^*}\big)$, i.e., cycles using only edges of $F^{\str A^*}$.

\medskip
\noindent [{\em $T$-cycle}] \ For every unary $T\in \big\{U_{\min}, U_{\max},
B, C\big\}$, a $T$-cycle is the cycle introduced for an $a\in T^{\str A}$.

\medskip
\noindent [{\em ear-cycle}] \ These are the cycles constructed as ears on the
gadgets for the relations $S^{\str A^*}$ and $L^{\str A^*}$.

\medskip
\noindent [{\em mixed-cycle}] \ All the other cycles are \emph{mixed}.

\medskip
\noindent For example, we get a mixed cycle if we start with $a_2$, $a'_0$,
$a_1$ in Figure~\ref{fig:relationF}
and then add the path introduced for $(a_1, a_2)\in S^{\str A}$ (ignoring the
ear).
\medskip

\noindent
A number of observations for these types of cycles are in order.
\begin{lem}\label{lem:GAcycles}
\begin{enumerate}
\item[(i)] All the $F$-cycles are of even length.\footnote{Moreover one can
    show that every \emph{chordless} $F$-cycle has length $4$.}

\item[(ii)] Every $U_{\min}$-, $U_{\max}$-, $B$-, and $C$-cycle is of
    length $5$, $7$, $9$, and $11$, respectively.

\item[(iii)] Every ear-cycle is of length $13$ or $15$.

\item[(iv)] 
    Every mixed-cycle neither uses new vertices of any $T$-cycle for $T\in
    \big\{U_{\min}, U_{\max}, B, C\big\}$ nor any vertex of any ear-cycle.

\item[(v)] Every mixed-cycle has length at least $17$.
\end{enumerate}
\end{lem}

\proof (i) follows easily from the fact that $\big(A^*,F^{\str A^*}\big)$ is
a bipartite graph; (ii) and (iii) are trivial.
\smallskip

\noindent For (iv)  assume that a mixed-cycle uses a \emph{new} vertex $b$ of
a $T$-cycle $\str C$ introduced for some $a\in T^{\str A^*}$, where $T\in
\big\{U_{\min}, U_{\max}, B, C\big\}$. As $\str C$ is mixed, it must contain
a vertex $c\notin T^{\str A^*}$. To reach $b$ from $c$ the mixed cycle must
pass through~$a$ and hence must contain one of the two segments of $\str C$
between $b$ and $a$. As a consequence, in order for the mixed-cycle to go
back from $b$ to $c$, it must also use the other segment of $\str C$
between~$a$ and $b$. This means that it must be the $T$-cycle $\str C$
itself, instead of a mixed one. A similar argument shows that mixed cycles do
not contain vertices of any ear-cycle.

\smallskip
\noindent To prove~(v), let $\str C$ be a mixed-cycle. By (iv), $\str C$ must
contain all vertices of a (at least one) path introduced for a pair
$(a,a')\in L^{\str A*}$ or $(a,b)\in S^{\str A^*}$ (ignoring the ear). As
this path has length $17$, we get our claim. \proofend

Conversely, given a $\tau_E$-structure $G$, which is a graph, we construct a
$\tau_0$-structure which we denote by $\str O(G)$, possibly the empty
structure. Recall the definitions of ``cycle'' and of ``path with ear'' given
by Lemma~\ref{lem:path-cycle}.
\begin{itemize}
\item $O(G):= \big\{(a_1,a_2)\in V(G)\times V(G) \bigmid \text{$a_1$ is a
    member of a cycle of length $9$, \ $a_2$ is a member} \\
    \text{of a cycle of length $11$, and there is a path from $a_1$ to
    $a_2$ of length $17$ with a $15$-ear}\big\}$

\item $<^{\str O(G)}:= \big\{((a_1,a_2), (b_1,b_2)) \in O(G)\times O(G)
    \bigmid \{a_2,b_1\}\in E(G)\big\}$

\item $U_\min^{\str O(G)}:= \big\{(a_1,a_2)\in O(G)\bigmid \text{$a_1$ is a
    member of a cycle of $5$ elements}\big\}$

\item $U_\max^{\str O(G)}:= \big\{(a_1,a_2) \in O(G)\bigmid \text{$a_1$ is
    a member of a cycle of $7$ elements}\big\}$

\item $S^{\str O(G)}:= \big\{((a_1,a_2),(b_1,b_2))\in O(G)\times O(G) \mid
    \text{there is a path from $a_1$ to $b_1$ of length $17$} \\
    \text{with a $13$-ear} \big\}$.

\end{itemize}

\begin{lem}\label{lem:fp}
For every complete $\tau_0$-ordering $\str A$ we have $\str O(G(\str A))\cong
\str A$.
\end{lem}

\proof Let $G:= G(\str A)$ and $\str A^+:= \str O(G)$. We claim that the
mapping $h: A\to A^+$ defined by
\[
h(a):= (a,a') \quad \text{for $a\in A$}
\]
is an isomorphism from $\str A$ to $\str A^+$. To that end, we first prove
that
\[
A^+= \big\{(a, a')\bigmid a\in A\big\},
\]
which implies that $h$ is well defined and a bijection. For every $a\in A$ it
is easy to see that $(a, a')\in O(G) \ (= A^+)$. For the converse, let $(a_1,
a_2)\in O(G)$. In particular, $a_1$ is a member of a cycle of length $9$. By
Lemma~\ref{lem:GAcycles}, this must be a $B$-cycle which contains some $a\in
A$. Using the same argument, $a_2$ is a member of a $C$-cycle which contains
a vertex $b'$ being the companion of some $b\in A$. Furthermore, there is a
path from $a_1$ to $a_2$ of length $17$ with a $15$-ear. The $15$-ear is a
cycle of length $15$. Again by Lemma~\ref{lem:GAcycles} this cycle is an
ear-cycle which belongs to the gadget we introduced for some $(c,c')\in
L^{\str A^*}$ with $c\in A$. Then it is easy to see that $a= c= b$. This
finishes the proof that $h$ is a bijection from $A$ to $A^+$.

Similarly, we can prove that $h$ preserves all the relations. \proofend

\noindent
We want to show that we can
obtain $\str O(G)$ from $G$ by a strongly existential \FO-interpretation.
We set
\begin{align*}
\eta(x,x',\bar x,\bar x',\bar z, \bar w)
 := &\ \text{``$\bar x$ is a cycle of length $9$ containing $x$,
  \ $\bar x'$ is a cycle of length $11$ containing $x'$}, \\
    & \quad \text{and $\bar z$ is a path from $x$ to $x'$ of length $17$ with the $15$-ear $\bar w$''}\\
  = &\ \varphi_{c9}(x,\bar x)\wedge\varphi_{c11}(x',\bar x')\wedge\varphi_{pe,17,15}(x,x'\bar z, \bar w).
\end{align*}
We define the desired interpretation $I$ of width $2$ of $\tau_0$-structures in
graphs. We set
\[
\varphi_\univ(x,x'):= \exists \bar x\exists \bar x'\exists \bar z\exists \bar w
 \, \eta(x,x',\bar x,\bar x',\bar z, \bar w).
\]
Hence for every graph $G$,
\[
O_I(G)=
 \big\{(a_1,a_2)\in V(G)\times V(G)
  \bigmid \text{$G\models \exists \bar x\exists \bar x'\exists \bar z\exists \bar w
   \, \eta(a_1,a_2,\bar x,\bar x',\bar z, \bar w)$}\big\}.
\]
Furthermore we define
\begin{itemize}
\item $\varphi_{U_\min}(x,x'):= \exists \bar z\, \varphi_{c5}(x,\bar z)$,

\item $\varphi_{U_\max}(x,x'):= \exists \bar z\, \varphi_{c7}(x,\bar z)$,

\item $\varphi_S(x,x',y,y'):= \exists \bar z \exists \bar w\; \text{``$\bar
    z$ is a path of length $17$ from $x$ to $y$ with a $13$-ear $\bar
    w$''}$ \\[1mm]
  {\hspace*{2.5cm}}$=\exists \bar z \exists \bar w
  \varphi_{pe,17,13}(x,\bar z,\bar w)$.
\end{itemize}

Then we have:
\begin{lem}\label{lem:fpnew}
The interpretation $I$ given by $\big(\varphi_\univ, \varphi_<,
\varphi_{U_\min}, \varphi_{U_\max}, \varphi_S\big)$ is strongly existential.
For every complete $\tau_0$-ordering $\str A$ we have $\str O_I(G(\str A))=
\str O(G(\str A))$ and hence, by Lemma~\ref{lem:fp},
\[
\str O_I(G(\str A))\cong \str A.
\]
\end{lem}

\noindent Setting $\psi:=\varphi_\Graph$, the sentence axiomatizing the class
of graphs, we get from Proposition~\ref{pro:ltfm}:

\begin{theo}[Tait's Theorem for graphs]\label{thm:ltfg}
There is a $\tau_E$-sentence $\varphi$ such that $\Graph_\fin(\varphi)$, the
class of finite graphs that are models of $\varphi$, is closed under induced
subgraphs but $\varphi$ is not equivalent to a universal sentence in finite
graphs.
\end{theo}

In this section we presented a strongly existential interpretation of
$\tau_0$-structures and applied it to finite complete $\tau_0$-orderings,
i.e, to models of $\varphi_0\wedge \varphi_1$. A straightforward
generalization of the preceding proofs allows to show the following result
for vocabularies obtained from $\tau_0$ by adding pairs. We shall use it in
Section~\ref{sec:gur}.

\begin{lem}\label{lem:fppairs}
Let $\tau$ be obtained from $\tau_0$ by adding pairs. There is a strongly
existential interpretation $I\ (=I_\tau)$ that for every extension
$\varphi_{0\tau}$ of $\varphi_0$ assigns to every $\tau$-structure $\str A$
that is a model of $\varphi_{0\tau}\wedge \varphi_{1\tau}$ a graph $G(\str
A)$ with $\str O_I(G(\str A))\cong \str A$. For finite $\str A$ the graph
$G(\str A)$ is finite.
\end{lem}

\proof We get the graph $G(\str A)$ as in the case $\tau:= \tau_0$: For the
elements of new unary relations we add cycles such that the lengths of the
cycles are odd and distinct for distinct unary relations in $\tau$. Let $c$
be the maximal length of these cycles. Then we add paths with ears to the
tuples of binary relations as above. For distinct binary relations the ears
should have distinct length and again this length should be odd and greater
than $c$. On the other hand, the length of added new paths can be the same
for all binary relations but should be greater than the length of all the
cycles. \proofend

\begin{rem}\label{rem:genpic}
(a) Let $\cls C:=\Mod_\fin(\forall x\neg Exx)$ be the class of directed
graphs. Then $\cls C':= \Graph_\fin$, the class of finite graphs, is a
subclass of $\cls C$ closed under induced substructures and definable in
$\cls C$ by the universal sentence $\forall x \forall y(Exy\to Eyx)$. As the
\LT\ Theorem fails for the class of finite graphs, it fails for the class of
directed graphs by Remark~\ref{rem:ltsub}.

\medskip
\noindent (b) Now let $\cls C:= \Graph_\fin$ and $\cls C':=
\textsc{Planar}_\fin$ be the class of finite planar graphs, a subclass of
$\Graph_\fin$ closed under induced subgraphs. As mentioned in the
Introduction, in~\cite{atsdawgro08} it is shown that the \LT\ Theorem fails
for $\textsc{Planar}_\fin$. As $\textsc{Planar}_\fin$ is not axiomatizable
in~$\Graph_\fin$ by a universal sentence, not even by a first-order sentence,
we do not get the failure of the \LT\ Theorem for the class of finite graphs,
i.e., Tait's Theorem for graphs, by applying the result of
Remark~\ref{rem:ltsub}. We show that $\textsc{Planar}_\fin= \Forb_\fin(\cls
F)$ for a finite set $ \cls F$ of finite graphs (or, equivalently,
$\textsc{Planar}_\fin= \Mod_\fin(\mu)$ for a universal $\mu$) leads to a
contradiction. Let $k$ be the maximum size of the set of vertices of graphs
in $\cls F$. Let $G$ be the graph obtained from the clique $K_5$ of 5
vertices by subdividing each edge by $k+1$. Clearly, $G\notin
\textsc{Planar}_\fin$. However, every subgraph of $G$ induced on at most $k$
elements is planar. Hence, $G\in \Forb_\fin(\cls F)$.

\medskip
\noindent (c) Let $\tau$ be any vocabulary with at least one at least binary
relation $T$. Then the \LT\ Theorem fails for the class $\cls
C:=\Str_\fin[\tau]$, the class of all finite $\tau$-structures. By
Remark~\ref{rem:ltsub} it suffices to show the existence of a universally
definable subclass $\cls C'$ of $\cls C$ which ``essentially is the class of
graphs.'' We set
\[
\mu:=\forall x\forall \bar u\neg Txx\bar u
 \wedge\forall x\forall y \forall\bar u\forall\bar v (Txy\bar u\to Tyx\bar v)
  \wedge\bigwedge_{R\in\tau,\ R\ne T}\forall \bar u \neg R\bar u
\]
and let $\cls C'$ be $\Mod_\fin(\mu)$.

If $\tau$ only contains unary relation symbols, the \LT\ Theorem holds for
$\Str_\fin[\tau]$. It is easy to see for an $\FO(\tau)$-sentence $\varphi$
that the closure under induced substructures of $\Mod_\fin(\varphi)$ implies
that of $\Mod(\varphi)$.
\end{rem}

\section{Gurevich's Theorem}\label{sec:gur}

The following discussion will eventually lead to a proof of Gurevich's
Theorem, i.e., Theorem~\ref{thm:igur}. Our proof essentially follows
Gurevich's proof in~\cite{gur84}, but it contains some elements of Rossman's
proof of the same result in~\cite{ros12}.~\footnote{The reader
of~\cite{gur84} will realize that the definition of $\varphi^n$ on page~190
of~\cite{gur84} must be modified in order to ensure that the class of models
of $\varphi^n$ is closed under induced substructures.} Afterwards we show
that it remains true if we restrict ourselves to graphs.

\medskip
\noindent Our main tool is Proposition~\ref{pro:agr}, and the goal is to
construct a formula $\gamma$ in~\eqref{eq:agr} whose size is much smaller
than the number $m$. Basically $\gamma$ will describe a very long computation
of a Turing machine on a short input. We fix a universal Turing machine $M$
operating on an one-way infinite tape, the tape alphabet is $\{0, 1\}$, where
$0$ is also considered as blank, and $Q$ is the set of states of $ M$. The
initial state is $q_0$, and $q_h$ is the halting state; thus $q_0, q_h\in Q$
and we assume that $q_0\ne q_h$. An instruction of $M$ has the form
\[
qapbd,
\]
where $q, p\in Q$, $a,b\in \{0, 1\}$ and $d\in \{-1, 0, 1\}$. It indicates
that if~$ M$ is in state $q$ and the head of~$ M$ reads an~$a$, then the head
replaces $a$ by $b$ and moves to the left (if $d= -1$), stays still (if $d=
0$), or moves to the right (if $d= 1$).
In order to describe computations of $M$ by \FO-formulas
we introduce binary predicates $H_q(x,t)$ for $q\in Q$ to indicate that at
time $t$ the machine is in state $q$ and the head scans cell $x$, and a
binary predicate $C_0(x,t)$ to indicate that the content of cell $x$ at time
$t$ is 0.

The vocabulary $\tau_M$ is obtained from $\tau_0$ by adding pairs \big(see
Definition~\ref{def:pair}\,(a)\big),
\[
\tau_M:= \tau_0\cup \big\{H_q, H_q^\comp \bigmid q\in Q\big\}
\cup \big\{C_0, C_0^\comp\big\}.
\]
Intuitively, $H_q^\comp(x,t)$ says that ``at time $t$ the machine is not in
state $q$ or the head is not in cell $x$;'' and $C_0^\comp(x,t)$ says that
``at time $t$ the content of cell $x$  is (not 0 and thus is) 1.'' Sometimes
we write $C_1$ instead of $C_0^\comp$ (e.g., below in $\varphi_2$ if $a= 1$
or $b= 0$).

Let $\varphi_0$ and $\varphi_1$ be the sentences already introduced in
Section~\ref{sec:idea}. For $w\in \{0, 1\}^*$ the sentence~$\varphi_{0w}$
will be an extension of $\varphi_0$ \big(compare
Definition~\ref{def:pair}\,(b)\big); hence, $\varphi_{0w}$ will be a
universal sentence and all relations symbols besides $<$ are negative in
$\varphi_{0w}$; in particular, it contains as conjuncts $\varphi_0$ and
\[
\forall x\forall t \big(\neg C_0(x,t)
 \vee \neg C^\comp_0(x,t)\big)
 \wedge \bigwedge_{q\in Q} \forall x\forall t
                                \big(\neg H_q(x,t)\vee \neg H_q^\comp(x,t)\big).
\]
Finally, $\varphi_{0w}$ will contain the following sentences $\varphi_2$ and
$\varphi_w$ as conjuncts. The sentence $\varphi_2$ describes one
computation step. It contains for each instruction of $M$ one conjunct. For
example, the instruction $qapb1$ contributes the conjunct
\begin{align*}
\forall x x'\forall t t' \forall y
 \Big(\big(H_{q}&(x,t) \wedge C_a (x,t) \wedge S(x,x')\wedge S(t,t')\big) \\
  \to \big( &(\neg C_{1-b}(x,t') \wedge \neg H^\comp _{p}(x',t')) \\
  & \wedge (y\ne x'\to \bigwedge_{r\in Q}\neg H_r(y,t')) \\
  & \wedge (y\ne x\to((C_0(y,t)\to \neg C^\comp_0(y,t))\wedge (C^\comp_0(y,t')\to\neg C_0(y,t'))))\big)\Big).
\end{align*}
For $w\in \{0,1 \}^*$ the sentence $\varphi_w$ describes the initial
configuration of $M$ with input $w$ \big(if $w= w_1\ldots w_{|w|}$, the first
$|w|$ cells contain $w_1, \ldots, w_{|w|}$, the remaining cells contain $0$,
and the head scans the first cell in the starting state $q_0$\big). Hence, as
$\varphi_w$ we can take the conjunction of
\begin{itemize}
\item $\forall x_1\ldots \forall x_{|w|}\big((U_\min\, x_1\wedge
    \bigwedge_{i\in[|w|-1]}S{x_i}x_{i+1}) \\[2mm]
    {\hspace{4cm}} \to(\bigwedge_{\substack{i\in[|w|], \\
    w_i=0}}\neg C^\comp_{0}(x_i,x_1)\wedge
    \bigwedge_{\substack{i\in[|w|],\\w_i=1}} \neg C_{0}(x_i,x_1))\big)$

\item $\forall x_1\ldots \forall x_{|w|}\forall x\big((U_\min\, x_1\wedge
    \bigwedge_{i\in[|w|-1]}S{x_i}x_{i+1}\wedge x_{|w|}<x)\to \neg
    C^\comp_{0}(x,x_1)\big)$

\item $\forall x\forall y\big(U_\min\, x\to (\neg H_{q_0}^\comp(x,x)\wedge
    (y\ne x\to \bigwedge_{q\in Q}\neg H_q(y,x)))\big)$.
\end{itemize}
Note that $U_\min$, $U_\max$, and $S$ are negative in $
\varphi_{0w}$. We set $\varphi_{1M}:=\varphi_{1\tau_M}$; recall that by
Definition~\ref{def:pair}\,(c),
\[
\varphi_{1M}=
 \varphi_1\wedge\forall x\forall t \big(C_0(x,t)\vee C^\comp_0(x,t)\big)
 \wedge \bigwedge_{q\in Q}\forall x\forall t \big(H_q(x,t)\vee H_q^\comp(x,t)\big).
\]
Let $w\in \{0,1\}^*$ and $r\in \mathbb N$. Furthermore, let $\str A$ be a
$\tau_M$-structure where $ <^{\str A}$ is an ordering and $|A|\ge r+1$. Let
$a_0, \ldots, a_r$ be the first $r+1$ elements of $<^{\str A}$. Assume that
$M$ on the input $w\in\{0,1 \}^*$ runs at least $ r$ steps. We say that
\emph{$\str A$ correctly encodes $r$ steps of the computation of $M$ on $w$}
if for $i,j$ with $0\le i,j\le r$,
\begin{eqnarray}\label{eq:canmod1}
(a_i,a_j)\in C_0^{\str A}
 & \iff & \text{the content of cell $i$ after $j$ steps is 0}
\end{eqnarray}
and for $q\in Q$,
\begin{eqnarray}\label{eq:canmod2}
(a_i,a_j)\in H_q^{\str A} & \iff &
\text{after $j$ steps $M$ is in state $q$ and the head scans cell $j$.}
\end{eqnarray}
From the definitions of the sentences $\varphi_{0w}$ and $\varphi_{1M}$, we see:

\begin{lem}\label{lem:maxs}
Let $w\in \{0, 1\}^*$ and $\str A$ be a model $\varphi_{0w}\wedge
\varphi_{1M}$. If for $r\in \mathbb N$ we have $r+1\le |A|$ (in particular,
if $A$ is infinite) and $M$ on $w$ runs at least $r$ steps, then $\str A$
correctly encodes $r$ steps of the computation of $M$ on $w$.
\end{lem}

\noindent Finally, let $\gamma_M$ be a sentence expressing that ``the machine
$M$ reaches the halting state $q_h$ in exactly `$\max$' steps,'' more
precisely,
\begin{equation}\label{eq:defgamm}\gamma_M:=
 \exists t\exists x
  \big(U_\max t \wedge H_{q_h}(x,t)\wedge
   \forall t'\forall y (t'< t\to \neg H_{q_h}(y,t'))\big).
\end{equation}
As a consequence of the preceding lemma, we obtain:
\begin{cor}\label{cor:maxs}
Let $w\in \{0,1\}^*$ and assume that $M$ on input $w$ eventually halts, say
in $h(w)$ steps, then
\[
\varphi_{0w}\wedge \varphi_{1M}\wedge \gamma_M
\]
has no infinite model but a model with exactly $h(w)+1$ elements (this model
is unique up to isomorphism).
\end{cor}

\proof Let $\str A\models \varphi_{0w}\wedge \varphi_{1M}\wedge \gamma_M$.
Then $\str A\upharpoonright\tau_0$ is a complete $\tau_0$-ordering and $\str
A$ contains the description of the complete halting computation of $M$ on the
input $w$. As the machine $M$ reaches the halting state in exactly $h(w)$
steps, we see that $|A|= h(w)+1$; in particular, $A$ is finite.

On the other hand, we can interpret~\eqref{eq:canmod1} and~\eqref{eq:canmod2}
as defining relations $C_0^{\str A}$ and $H_q^{\str A}$ on the set $A:=
\big\{a_0, \ldots, a_{h(w)}\big\}$ equipped with the ``natural'' ordering and
its corresponding relations $U_{\min}$, $U_{\max}$, and $S$. If furthermore
we let $(C^\comp_0)^{\str A}$ and $(H_q^\comp)^{\str A}$ be the complements
in $A\times A$ of $C_0^{\str A}$ and~$H_q^{\str A}$, respectively, we get a
model of $\varphi_{0w}\wedge \varphi_{1M}\wedge \gamma_M$ with exactly
$h(w)+1$ elements. \proofend

\noindent
We set
\begin{equation}\label{eq:defchiw}
\chi_w:=\varphi_{0w}\wedge (\varphi_{1M}\to\neg \gamma_M).
\end{equation}
By Proposition~\ref{pro:agr} and Corollary~\ref{cor:maxs}, we get:
\begin{lem}\label{lem:mwsub}
Let $M$ on input $w$ eventually halt, say in $h(w)$ steps. Then:
\begin{enumerate}
\item[(a)] $\Mod(\chi_w)$ is closed under $<$-substructures.

\item[(b)] If $\chi_w$ is finitely equivalent to a universal sentence
    $\mu$, then $|\mu| \ge h(w)+1$.
\end{enumerate}
\end{lem}

\noindent
Now we show the following version of Gurevich's Theorem.
\begin{theo}\label{thm:gr}
Let $f: \mathbb N\to \mathbb N$ be a computable function. Then there is a
$w\in\{0,1\}^*$ such that $\Mod(\chi_w)$ is closed under $<$-substructures
but $\chi_w$ is not finitely equivalent to a universal sentence of length
less than $f(|\chi_w|)$.
\end{theo}

\proof By the previous lemma it suffices to find a $w\in\{0,1\}^*$ such that
$M$ on input $w$ halts in $h(w)$ steps with
\[
h(w)\ge f( |\chi_w|).
\]
W.l.o.g.\ we assume that $f$ is increasing. An analysis of the formula
$\chi_{w}$ shows that for some $c_M\in \mathbb N$ we have for all
$w\in\{0,1\}^*$,
\begin{equation}\label{eq:cMl}
|\chi_w|\le c_M\cdot |w|.
\end{equation}
We define $g:\mathbb N\to \mathbb N$ by
\begin{equation}\label{eq:hig}
g(k):= f(5\cdot c_M\cdot k).
\end{equation}
Let $M_0$ be a Turing machine computing $g$, more precisely, the function
$1^k\mapsto 1^{g(k)}$. We code $M_0$ and~$ 1^k$ by a $\{0,1 \}$-string
$\textit{code}(M_0,1^k)$ such that $M$ on $\textit{code}(M_0,1^k)$ simulates
the computation of~$M_0$ on $1^k$.

Choose the least~$k$ such that for $w:= \textit{code}(M_0,1^k)$ we have
\begin{equation}\label{eq:lcd}
|w|\le 5k.
\end{equation}
The universal Turing machine $M$ on input $w$ computes $1^{g(k)}$ and thus
runs at least $g(k)$ steps, say, exactly $h(w)$ steps. By \eqref{eq:cMl} --
\eqref{eq:lcd}
\[
h(w)\ge g(k)= f(5\cdot c_M\cdot k)\ge f(c_M\cdot |w|)\ge f(|\chi_w|).
 \benda
\]

\noindent Finally we prove Gurevich's Theorem for graphs. For $\tau:= \tau_M$
let $I$ be an interpretation according to Lemma~\ref{lem:fppairs}. For $w\in
\{0, 1\}^*$ we consider the sentence
\begin{equation}\label{eq:defrho}
\rho_w:= \forall \bar x\neg \varphi_\univ(\bar x)
 \vee (\varphi_{0w}\wedge (\varphi_{1M}\to \neg \gamma_M))^I
  = \forall \bar x\neg \varphi_\univ(\bar x)\vee \chi_w^I.
\end{equation}
 That is, for $G\models \rho_w$, either the graph $G$
interprets an empty $\tau_M$-structure, or a $\tau_M$-structure which is a
model of $\chi_w$.
If $M$ halts in $h(w)$ steps on input $w$, then $\varphi_{0w}\wedge
\varphi_{1M}\wedge \gamma_M$ has no infinite model but a finite model with
$h(w)+1$ elements by Corollary~\ref{cor:maxs}. Hence taking in
Proposition~\ref{pro:grfm} as $\psi$ the sentence $\psi_\Graph$ axiomatizing
the class of graphs we get the following analogue of Lemma~\ref{lem:mwsub}.

\begin{lem}\label{lem:mwsubg}
Let $M$ on input $w$ halt in $h(w)$ steps. Then:
\begin{enumerate}
\item[(a)] $\Graph(\rho_w)$, the class of graphs that are model of
    $\rho_w$, is closed under induced subgraphs (and hence equivalent in
    the class of graphs to a universal sentence).

\item[(b)] If $\rho_w$ is equivalent in the class of finite graphs to a
    universal sentence $\mu$, then $|\mu|^2 \ge h(w)$.
\end{enumerate}
\end{lem}


\medskip
\begin{theo}[Gurevich's Theorem for graphs]\label{thm:grg}
Let $f:\mathbb N\to \mathbb N$ be a computable function. Furthermore, let
$\rho_w$ be defined by~\eqref{eq:defrho}, where $I$ is an interpretation for
$\tau:= \tau_M$ according to Lemma~\ref{lem:fppairs}. Then there is a
$w\in\{0,1\}^*$ such that $\Graph(\rho_w)$ is closed under induced subgraphs
but $\rho_w$ is not equivalent in the class of finite graphs to a universal
sentence of length less than $f(|\rho_w|)$.
\end{theo}

\proof Again we assume that $f$ is increasing. By the previous lemma it
suffices to find a $w\in\{0,1\}^*$ such that $M$ on input $w$ halts in $h(w)$
steps with
\[
h(w)\ge f( |\rho_w|)^2.
\]
There is a $c\in \mathbb N$, which depends on $I$ but not on $w$, such that
for $c_I$ as in~\eqref{eq:cci} and $c_M$ as in~\eqref{eq:cMl} we have for
$d_M:= c+ c_I\cdot c_M$,
\begin{equation}\label{eq:cMlg}
|\rho_w|\le c+ c_I \cdot |\chi_w|\le c+ c_I\cdot c_M \cdot |w|\le d_M\cdot |w|.
\end{equation}
We define $g:\mathbb N\to \mathbb N$ by
\begin{equation}\label{eq:higg} g(k):= f(5\cdot d_M\cdot k)^2
\end{equation}
and then proceed as in the proof of Theorem~\ref{thm:gr}. Let $M_0$ be a
Turing machine computing the function $1^k\mapsto 1^{g(k)}$. We code $M_0$
and~$ 1^k$ by a $\{0,1 \}$-string $\textit{code}(M_0,1^k)$ such that $M$ on
$\textit{code}(M_0,1^k)$ simulates the computation of $M_0$ on $1^k$.

Choose the least~$k$ such that for $w:= \textit{code}(M_0,1^k)$ we have
\begin{equation}\label{eq:lcdg}
|w|\le 5k.
\end{equation}
The universal Turing machine $M$ on input $w$ computes $1^{g(k)}$ and thus
runs at least $g(k)$ steps, say, exactly $h(w)$ steps. We have
\[
h(w)\ge g(k)= f(5\cdot d_M\cdot k)^2\ge f(d_M\cdot |w|)^2\ge f( |\rho_w|)^2
\]
by~\eqref{eq:cMlg} -- \eqref{eq:lcdg}. \proofend

\begin{rem}\label{rem:gur}
Using previous remarks (Remark~\ref{rem:sigma1m} and
Remark~\ref{rem:sigma1n}) one can even show that for every computable
function $f: \mathbb N\to \mathbb N$ the sentence $\chi_w$ is not finitely
equivalent to a $\Pi_2$-sentence of length less than $f(|\chi_w|)$ and the
sentence $\rho_w$ is not finitely equivalent in graphs to a $\Pi_2$-sentence
of length less than $f(|\chi_w|)$. Moreover, $\chi_w$ and $\rho_w$ are
equivalent to $\Sigma_2$.

For this purpose note that in models of $\varphi_{0w}$ the sentence
$\gamma_M$ is equivalent to
\[
\exists t\exists x
  \big(U_\max t \wedge H_{q_h}(x,t)\big)
   \wedge \forall t_1\forall t_2\forall y \big(t_1<t_2\to \neg H_{q_h}(y,t_2)\big).
\]
and hence equivalent to a $\Sigma_2$ and to a $\Pi_2$-sentence. One easily
verifies that the same holds for $\gamma_M^I$.

\end{rem}
\medskip

\section{Some undecidable problems}\label{sec:und}

In this section we show that various problems related to the results of the
preceding sections are undecidable. Among others, these results explain why
it might be hard, in fact impossible in general, to obtain forbidden induced
subgraphs for various classes of graphs.

\begin{prop}\label{pro:adcig}
There is no algorithm that applied to any $\FO[\tau_E]$-sentence $\varphi$
decides whether the class $\Graph(\varphi)$ is closed under induced
subgraphs.
\end{prop}

\proof Assume $\mathbb A$ is such an algorithm. By the Completeness Theorem
there is an algorithm $\mathbb B$ that assigns to every sentence $\varphi$
with $\Graph(\varphi)$ closed under induced subgraphs a universal sentence
equivalent to $\varphi$ in graphs. Define the function $g$ by
\[
g(\varphi):=
 \begin{cases}
 0, & \text{if $\mathbb A$ rejects $\varphi$} \\
 m, & \text{$\mathbb B$ needs $m$ steps to produce
            a universal sentence equivalent to $\varphi$}
 \end{cases}
\]
and set $f(k):= \max\{g(\varphi)\mid |\varphi|\le k\}$. Then $f$ would
contradict Gurevich's Theorem for graphs, i.e., Theorem~\ref{thm:grg}.
 \proofend

\begin{cor}
There is no algorithm that applied to any $\FO[\tau_E]$-sentence $\varphi$
either reports that $\Graph(\varphi)$ is not closed under induced subgraphs
or it computes for $\Graph(\varphi)$ a class of forbidden induced subgraphs.
\end{cor}

\proof Otherwise we could use this algorithm as a decision algorithm for the
previous result. \proofend

The following proposition is the analog of Proposition~\ref{pro:adcig} for
classes of finite graphs. We state it for $\FO[\tau_E]$-sentences and graphs
even though we prove it for $\FO[\tau_M]$-sentences. One gets the version for
graphs using the machinery we developed in previous sections similarly as we
do it to get Corollary~\ref{cor:cunif} from Proposition~\ref{pro:cunif}
below.

\medskip
We write $M: w\mapsto \infty$ for the universal Turing machine $M$ and a word
$w\in\{0,1 \}^*$ if $M$ on input~$ w$ does not halt. We make use of the
sentences $\varphi_{0w}$, $\varphi_{1M}$, and $\gamma_M$ defined in the
previous section.

\begin{prop}\label{pro:cisf}
There is no algorithm that applied to any $\FO[\tau_E]$-sentence $\varphi$
decides whether the class $\Graph\,_\fin(\varphi)$ is closed under induced subgraphs.
\end{prop}

\proof For the universal Turing machine $M$ and a word $w\in\{0,1 \}^*$
consider the sentence
\[
\pi_w:=\varphi_{0w}\wedge \varphi_{1M}\wedge
\gamma_M.
\]
Then
\begin{eqnarray}\label{eq:cisf}
\text{$\Mod\,_\fin(\pi_w)$ is closed under induced subgraphs}
 & \iff &
M: w\mapsto\infty.
\end{eqnarray}
In fact, if $M: w\mapsto \infty$, then $\Mod\,_\fin(\pi_w)= \emptyset$, hence
$\Mod\,_\fin(\pi_w)$ is trivially closed under induced subgraphs. If $M$ on
input $w$ halts after $h(w)$ steps, then, up to isomorphism, there is a
unique model~$\str A_w$ of $\pi_w$ and it has $h(w)+ 1$ elements. By
Lemma~\ref{lem:infm} every proper induced substructure of~$\str A_w$ is not a
model of~$\pi_w$. Hence $\Mod\,_\fin(\pi_w)$ is not closed under induced
subgraphs. As the halting problem for every universal Turing machine is not
decidable, by~\eqref{eq:cisf} we get our claim. \proofend

\begin{prop}\label{pro:cunif}
There is no algorithm that applied to any $\FO[\tau_M]$-sentence, which is
finitely equivalent to a universal sentence, computes such a universal
sentence.
\end{prop}

\proof Again we show that such an algorithm would allow us to decide for
every $w\in \{0, 1\}^*$ whether the universal Turing machine $M$ halts on
input $w$. In~\eqref{eq:defchiw} we defined $\chi_w$ by
\[
\chi_w= \varphi_{0w}\wedge (\varphi_{1M}\to \neg \gamma_M).
\]
If $M$ halts on $w$, by Lemma~\ref{lem:mwsub} we know that $\Mod(\chi_w)$ is
closed under $<$-substructures and thus equivalent to a universal sentence.
The claimed algorithm (or, even the Completeness Theorem) will produce such a
universal $\mu$. Furthermore, by Corollary~\ref{cor:maxs} we know that there
is a finite model with $h(w)+1$ elements, which is a model of
$\varphi_{0w}\wedge \neg\chi_w$, hence it is a model of $\varphi_{0w}\wedge
\neg \mu$.

If $M$ does not halt on $w$, then we show that $\Mod\,_\fin(\chi_w)=
\Mod\,_\fin(\varphi_{0w})$. Clearly $\Mod_\fin(\chi_w)\subseteq
\Mod_\fin(\varphi_{0w})$. Now let $\str A$ be a finite model of $\varphi_{0w}
$. If $\str A\not \models \varphi_{1M}$, then $\str A\models \chi_w$.
Otherwise $\str A \models \varphi_{1M}$, then $\str A$
correctly represents the first $|A|-1$ steps of the computation of $M$ on $w$
by Lemma~\ref{lem:maxs}. Thus~$\str A$ is
a model of $\neg \gamma_M$ as $M$ does not halt on $w$. Therefore, $\str A$
is a model of $\chi_w$.

Now we can see whether $M$ does not halt on $w$ by checking whether the
universal sentence produced by the claimed algorithm is finitely equivalent
to the universal sentence $\varphi_{0w}$. This can be checked effectively by
Corollary~\ref{cor:unileq} and Corollary~\ref{cor:unilfeq}. \proofend

\begin{cor}\label{cor:cunif}
There is no algorithm that applied to any $\FO[\tau_E]$-sentence $\varphi$
such that $\Graph\,_\fin(\varphi)$ has a finite set of forbidden induced
subgraphs computes such a set.
\end{cor}

\proof Equivalently we show that there is no algorithm that applied to any
$\FO[\tau_E]$-sentence~$\varphi$ such that $\Graph\,_\fin(\varphi)=
\Graph\,_\fin(\mu)$ for some universal sentence $\mu$ computes such a $\mu$.

For graphs let $I\ (=I_{\tau_M})$ be a strongly existential interpretation of
$\tau_M$-structures in graphs according to Lemma~\ref{lem:fppairs}. We know
that for every finite $\tau_M$-structure $\str A$ there is a finite graph $G$
such that $G_I\cong \str A$.


For $w\in\{0,1 \}^*$ we consider the sentence $\rho_w$ defined
in~\eqref{eq:defrho} in the proof of Theorem~\ref{thm:gr},
\[
\rho_w=
 \forall \bar x
  \neg \varphi_\univ(\bar x)\vee (\varphi_{0w}\wedge (\varphi_{1M}\to \neg \gamma_M))^I
  = \forall \bar x\neg \varphi_\univ(\bar x)\vee \chi_w^I.
\]
We show that $\rho_w$ is equivalent to a universal sentence $\mu$ on finite
graphs. Moreover, $M$ does not halt on input $w$ if and only if $\mu$ is
finitely equivalent to the universal sentence $\forall x\neg
\varphi_\univ(\bar x)\vee \varphi_{0w}^I$

\medskip
If $M$ halts on $w$, then $\varphi_{0w}\wedge \varphi_{1M}\wedge \gamma_M$
has no infinite model but a finite model $\str A$. Hence,
by~Proposition~\ref{pro:grfm} we know that $\Graph(\rho_w)$ is closed under
induced subgraphs. Therefore, $\rho_w$ is equivalent to a universal sentence
$\mu$ in $\Graph$. Let $G$ be a finite graph with $G_I\cong \str A$. Then
$G\models (\varphi_{0w}\wedge \varphi_{1M}\wedge \gamma_M)^I$ and thus,
$G\models \neg \rho_w$. Hence $G$ is a finite graph which is a model of
$\varphi^I_{0w}\wedge \neg \mu$. This means that $\mu$ is not equivalent to
$\forall x\neg \varphi_\univ(\bar x)\vee \varphi_{0w}^I$ on all finite
graphs, as $G$ is also a model of $\forall x\neg \varphi_\univ(\bar
x)\vee\varphi_{0w}^I$.

If $M: w\to \infty$, then we show that $\Graph\,_\fin(\rho_w)=
\Graph\,_\fin(\forall x\neg \varphi_\univ(\bar x)\vee\varphi_{0w}^I)$.
Clearly $\Graph_\fin(\rho_w)\subseteq \Graph_\fin(\forall x\neg
\varphi_\univ(\bar x)\vee\varphi_{0w}^I)$. Now let the graph $G$ be a model
of $\forall x\neg \varphi_\univ(\bar x)\vee\varphi_{0w}^I$. Further we can
assume  that $G\models \exists x \varphi_\univ(\bar x)$. In particular, $\str
A:= G_I$ is well defined. If $G\not\models \varphi^I_{1M}$, then $G\models
\chi^I_w$ and therefore, $G\models \rho_w$. If $G\models \varphi^I_{1M}$,
then $\str A\models \varphi_{0w}\wedge\varphi^I_{1M}$. As $M: w\to \infty$,
by Lemma~\ref{lem:maxs} the structure $\str A$ correctly represents the first
$|A|- 1$ steps of the computation of $M$ on $w$. Thus, $\str A$ is a model of
$\neg \gamma_M$, again as $M$ does not halt on input~$w$. It follows that $G$
is a model of $\neg \gamma_M^I$, and then $G\models \rho_w$.

Now we can decide the halting problem for $M$. Given a word $w$, we use the
claimed algorithm to get a universal sentence $\mu$ equivalent to $\rho_w$ in the
class of graphs. Finally we check whether $\mu$ is
finitely equivalent to $\forall x\neg \varphi_\univ(\bar x)\vee
\varphi_{0w}^I$. This can be checked effectively again by
Corollary~\ref{cor:unileq} and Corollary~\ref{cor:unilfeq}. \proofend

Observe that Corollary~\ref{cor:cunif} is precisely
Theorem~\ref{thm:finiteuniversalgraph} as stated in the Introduction. Finally
we prove Theorem~\ref{thm:decisionfinuniv}, which is equivalent to the
following result.


\begin{theo}
There is no algorithm that applied to an $\FO[\tau_E]$-sentence $\varphi$
such that $\Graph_\fin(\varphi)$ is closed under induced subgraphs decides
whether there is a finite set $\cls F$ of graphs such that
\[
\Graph_\fin(\varphi)=\Forb_\fin(\cls F).
\]
\end{theo}

\proof Again we prove the corresponding result for $\tau_M$-sentences and
$\tau_M$-structures and leave it to the reader to translate it to graphs as
in the previous proof. That is, we show:

\begin{quote}
{\em There is no algorithm that applied to an $\FO[\tau_M]$-sentence
$\varphi$ such that $\Mod_\fin(\varphi)$ is closed under induced
substructures decides whether there is a finite set $F$ of finite
$\tau_M$-structures such that
\[
\Mod_\fin(\varphi)=\Forb_\fin(\cls F).
\]}
\end{quote}
For $w\in\{0,1 \}^*$ let
\[
\alpha_w:=\varphi_{0w}\wedge (\varphi_{1M}\to \gamma_M).
\]
We show that $\Mod_\fin(\alpha_w)$ is closed under induced subgraphs and that
\begin{eqnarray*}
M: w\to \infty
 & \iff &
\text{$\alpha_w$ is not finitely equivalent to a universal sentence}.
\end{eqnarray*}
Assume first that $M: w\to \infty$. Then $\varphi_{0w}\wedge
\varphi_{1M}\wedge \gamma_M$ has no finite model by~Lemma~\ref{lem:maxs} and
the definition~\eqref{eq:defgamm} of $\gamma_M$. Therefore,
$\Mod_\fin(\alpha_w)= \Mod_\fin(\varphi_{0w}\wedge \neg \varphi_{1M})$. By
Lemma~\ref{lem:exttai} we know that $\Mod_\fin(\varphi_{0w}\wedge \neg
\varphi_{1M})$ is closed under induced substructures but not finitely
equivalent to a universal sentence.

Now assume that $M$ on input $w$ halts in $h(w)$ steps. Then
Corollary~\ref{cor:maxs} guarantees that there is a unique model $\str A_w$
of $\varphi_{0w}\wedge \varphi_{1M}\wedge \gamma_M$ with $|A_w|= h(w)+1$. We
present a finite set $\cls F$ of finite $\tau_M$-structures such that
\begin{equation}\label{eq::prfis}
\Mod_\fin(\alpha_w)= \Forb_\fin(\cls F).
\end{equation}
As $\varphi_{0w}$ is universal, there is a finite set $\cls F_0$ of finite
$\tau_M$-structures such that
\[
\Mod_\fin(\varphi_{0w})= \Forb_\fin(\cls F_0).
\]
Moveover, we set
\[
\cls F_1:= \big\{\str B\in \Str[\tau_M]
  \bigmid \str B\models \varphi_{0w}\wedge\varphi_{1M}
   \text{\ and $B= [\ell]$ for some $\ell\le h(w)$}\big\}
\]
and
\begin{align*}
\cls F_2:= \big\{\str B\in\Str[\tau_M]
 \bigmid \str B\models \varphi_{0w}\wedge \varphi^*_{1M}
  \wedge \forall t \forall t'(t< t'\to \forall y &\neg H_{q_h}(y,t)) \\
  & \text{and $B= [h(w)+2]$}\big\}.
\end{align*}
Here $\varphi^*_{1M} $is obtained from $\varphi_{1M}$ by replacing the
conjunct $\varphi_1$ \big(see~\eqref{eq:var1}\big) by
\[
\varphi^*_1:= \exists x U_\min x\wedge \forall x\forall y(x<y\to\exists z Sxz).
\]
The difference is that $\varphi^*_1$ does not require the set $U_{\max}$ to
be nonempty. Hence,
\[
\varphi^*_{1M} =
 \varphi^*_1\wedge\forall x\forall t \big(C_0(x,t)\vee C^\comp_0(x,t)\big)
 \wedge \bigwedge_{q\in Q}\forall x\forall t \big(H_q(x,t)\vee H_q^\comp(x,t)\big).
\]
Note that Lemma~\ref{lem:maxs} remains true if in its statement we replace
$\varphi_{1M}$ by $\varphi^*_{1M}$.

For $\cls F:= \cls F_0\cup \cls F_1\cup \cls F_2$ we show \eqref{eq::prfis}.
Assume first that a finite structure $\str C$ is a model of $\alpha_w$. In
particular, $\str C\models \varphi_{0w}$ and therefore, $\str C$ has no
induced substructure isomorphic to a structure in $\cls F_0$.

Now, for a contradiction suppose that $\str B$ is an induced substructure of
$\str C$ isomorphic to a structure in~$\cls F_1$. Then $\str B\models
\varphi_{1M}$ and thus, by Lemma~\ref{lem:infm}, $\str C= \str B$. As $\str
C\models \alpha_w$, we get $\str C\models \varphi_{0w}\wedge
\varphi_{1M}\wedge \gamma_M$. Hence, $\str C\cong \str A_w$, a contradiction,
as on the one hand $|C|= |B|\le h(w)$ and on the other hand $|C|= |A_w|=
h(w)+1$.

Next we show that $\str C$ has no induced substructure $\str B$ isomorphic to
a structure in $\cls F_2$. As $\str B\models \varphi_{0w}\wedge
\varphi^*_{1M}$ and has $h(w)+2$ elements, the first $h(w)+1$ elements of
$\str B$ correctly encode the first $h(w)$ steps of the computation of $M$ on
$w$, hence the full computation. As $|B|= h(w)+2$, this contradicts $\str
B\models \forall t \forall t' \big(t<t'\to \forall y\neg H_{q_h}(y,t)\big)$.

\medskip
\noindent As the final step let $\str C\in \Forb_\fin(\cls
F)$. We show that $\str C\models \alpha_w$. As $\str C$ omits the structures
in $\cls F_0$ as induced substructures, we see that $\str C\models
\varphi_{0w}$. If $\str C\not\models \varphi_{1M}$, we are done.

Recall that by Lemma~\ref{lem:maxs} (more precisely, by the extension of
Lemma~\ref{lem:maxs} mentioned above) for finite structures $\str B$ of
$\varphi_{0w}\wedge \varphi^*_{1M}$ we know:
\begin{itemize}
\item[(a)] if $|B|\le h(w)+1$, then $\str B$ encodes $|B|- 1$ steps of the
    computation of $M$ on $w$,

\item[(b)] if $|B|> h(w)+1$, then the first $h(w)+1$ elements in the
    ordering $<^{\str B}$ correctly encode the (full) computation of $M$ on
    $w$.
\end{itemize}
Now assume that $\str C\models \varphi_{1M}$, then (a) and (b) apply to $\str
C$. As no structure in $\cls F_1$ is isomorphic to an induced substructure of
$\str C$, we see that $|C|\ge h(w)+1$. But $\str C$ cannot have more than
$h(w)+1$ elements, as otherwise the substructure of $\str C$ induced on the
first $h(w)+2$ elements would be isomorphic to a structure \str B in $F_2$, a
contradiction. \proofend

\begin{rem}\label{rem:und}
Mainly using Remark~\ref{rem:gur} one easily verifies that in all results
but Proposition~\ref{pro:cisf} of this section we can replace
\[
\text{There is no algorithm that applied to an $\FO[\tau_E]$-sentence $\varphi$ \ldots}
\]
by
\[
\text{There is no algorithm that applied to a $\Sigma_2$-sentence $\varphi$ \ldots}
\]
In Proposition~\ref{pro:cisf} we have to replace it by
\[
\text{There is no algorithm that applied to a $\Pi_2$-sentence $\varphi$ \ldots}
\]
as $\varphi_{1M}$ (and $\varphi^I_{1M}$) are $\Pi_2$-sentences.
\end{rem}

\bibliographystyle{plain}
\bibliography{refer}

\end{document}

%% file: pathear.tex
\begin{tikzpicture}[
  scale=1.5,
  vertex/.style={circle,inner sep=0pt,minimum size=1.7mm,fill=black},
  ]

  \begin{scope}
    \foreach \x in {0,1,...,6}
           \node[vertex] (u\x) at (\x,1) {};


   \path (u0) node[above] {\footnotesize $a$}
             (u6) node[above] {\footnotesize $b$};

   \foreach \x in {0,1,...,5}
          \draw[-] (\x+0.1,1)--(\x+0.9,1);

   \node[vertex] (e0) at (5,1.5) {};
   \node[vertex] (e1) at (4.5,2) {};
   \node[vertex] (e2) at (5.5,2) {};
   \node[vertex] (e3) at (5,2.5) {};

   \draw[-] (5,1.07)--(5,1.43);

   \draw[-] (4.95,1.55)--(4.55,1.95);
   \draw[-] (5.05,1.55)--(5.45,1.95);

   \draw[-] (4.55,2.05)--(4.95,2.45);
   \draw[-] (5.45,2.05)--(5.05,2.45);

  \end{scope}

\end{tikzpicture}

%% file: relationF.tex
\begin{tikzpicture}[
  scale=1.2,
  vertex/.style={circle,inner sep=0pt,minimum size=1.7mm,fill=black},
  ]

\begin{scope}
    \foreach \x in {0,1,...,5}
           \node[vertex] (a\x) at (\x,1) {};

    \foreach \x in {0,1,...,5}
           \path (a\x) node[below] {\footnotesize $a_{\x}$};

   \foreach \x in {0,1,...,4}
          \draw[->] (\x,1)--(\x+.95,1);

   \foreach \x in {0,1,...,3}
      {
      \pgfmathtruncatemacro{\xplusone}{\x + 1};
       \foreach \y in {2,...,5}
          \ifthenelse{\y>\xplusone}{\draw[->] (\x,1)..controls (\x+.1,.8+\y*.3+\x*.1) and (\y-.1,.8+\y*.3+\x*.1)..(\y,1.05);}{};
      }

\end{scope}

\begin{scope}[xshift=7cm,yshift=0cm]

    \foreach \x in {0,1,...,5}
           \node[vertex] (a\x) at (\x,1) {};

    \foreach \x in {0,1,...,5}
           \path (a\x) node[below] {\footnotesize $a_{\x}$};

    \foreach \x in {0,1,...,5}
           \node[vertex] (ap\x) at (\x,2.2) {};

    \foreach \x in {0,1,...,5}
           \path (ap\x) node[above] {\footnotesize $a'_{\x}$};

   \foreach \x in {0,1,...,5}
      {
       \foreach \y in {0,1,...,5}
          \ifthenelse{\y<\x}{\draw[-] (\y,2.15)--(\x,1.05);}{};
      }

\end{scope}

\end{tikzpicture}